\documentclass[aps,prd,superscriptaddress,amsfonts,amssymb,amsmath,showpacs,10pt,onehalfspacing]{revtex4-2}
\usepackage[utf8]{inputenc}
\usepackage{times}
\usepackage{siunitx}
\usepackage{url}
\usepackage{tensor}
\usepackage{longtable}
\usepackage{mathtools}
\usepackage{enumitem}
\usepackage{amssymb}
\usepackage{amsmath}
\usepackage{makecell,tabularx,multirow}
\usepackage{color, colortbl}
\usepackage{adjustbox}
\usepackage{tikz}
\usetikzlibrary{positioning}
\usetikzlibrary{shapes.geometric, arrows}
\usepackage[pages=some]{background}
\usepackage{hyperref}
\usepackage{nameref}
\usepackage{accents}
\usepackage{todonotes}
\usepackage{eqnarray}
\usepackage{makecell}
\def\layersep{2.5cm}
\usepackage{makecell}
\hypersetup{
    colorlinks=true,
    linkcolor=blue,
    filecolor=blue,
    urlcolor=blue,
    citecolor=blue
}
\usepackage{comment}
\usepackage{booktabs}

\newcolumntype{C}{>{\centering\arraybackslash}X}
\newcolumntype{x}[1]{>{\centering\arraybackslash\hspace{0pt}}p{#1}}

\renewcommand{\arraystretch}{1.5}

\tikzstyle{startstop} = [rectangle, rounded corners, 
text width=7em, minimum height=1cm,text centered, draw=black, fill=red!30]
\tikzstyle{io} = [trapezium, trapezium left angle=70, trapezium right angle=110, 
text width=7em, minimum height=1cm, text centered, draw=black, fill=blue!30]
\tikzstyle{process} = [rectangle, 
text width=7em, minimum height=1cm, text centered, draw=black, fill=orange!30]
\tikzstyle{decision} = [diamond, minimum height=1cm, text centered, text width=5.5em, node distance=3cm, draw=black, fill=green!30]
\tikzstyle{arrow} = [thick,->,>=stealth]

\begin{document}

\title{Model-independent calibration of Gamma-Ray Bursts with neural networks}

\author{Purba Mukherjee}
\email{pdf.pmukherjee@jmi.ac.in}
\affiliation{Centre for Theoretical Physics, Jamia Millia Islamia, New Delhi - 110025, India}

\author{Maria Giovanna Dainotti}
\email{maria.dainotti@nao.ac.jp }
\affiliation{Division of Science, National Astronomical Observatory of Japan, 2-21-1 Osawa, Mitaka, 181-8588 Tokyo, Japan}
\affiliation{The Graduate University for Advanced Studies (SOKENDAI), Shonankokusaimura, Hayama, Miura District, Kanagawa 240-0115, Japan}
\affiliation{Space Science Institute, 4765 Walnut St Ste B, Boulder, CO 80301, USA}
\affiliation{Nevada Center for Astrophysics, University of Nevada, 4505 Maryland Parkway, Las Vegas, NV 89154, USA}

\author{Konstantinos F. Dialektopoulos}
\email{kdialekt@gmail.com}
\affiliation{Department of Mathematics and Computer Science, Transilvania University of Brasov, Eroilor 29, Brasov, Romania}

\author{Jackson Levi Said}
\email{jackson.said@um.edu.mt}
\affiliation{Institute of Space Sciences and Astronomy, University of Malta, Msida, Malta}
\affiliation{Department of Physics, University of Malta, Msida, Malta}

\author{Jurgen Mifsud}
\email{jurgen.mifsud@um.edu.mt}
\affiliation{Institute of Space Sciences and Astronomy, University of Malta, Msida, Malta}
\affiliation{Department of Physics, University of Malta, Msida, Malta}

\begin{abstract}
The $\Lambda$ Cold Dark Matter ($\Lambda$CDM) cosmological model has been highly successful in predicting cosmic structure and evolution, yet recent precision measurements have highlighted discrepancies, especially in the Hubble constant inferred from local and early-Universe data. Gamma-ray bursts (GRBs) present a promising alternative for cosmological measurements, capable of reaching higher redshifts than traditional distance indicators. This work leverages GRBs to refine cosmological parameters independently of the $\Lambda$CDM framework. Using the Platinum compilation of long GRBs, we calibrate the Dainotti relations—empirical correlations among GRB luminosity properties—as standard candles through artificial neural networks (ANNs). We analyze both the 2D and 3D Dainotti calibration relations, leveraging an ANN-driven Markov Chain Monte Carlo approach to minimize scatter in the calibration parameters, thereby achieving a stable Hubble diagram. This ANN-based calibration approach offers advantages over Gaussian processes, avoiding issues such as kernel function dependence and overfitting. Our results emphasize the need for model-independent calibration approaches to address systematic challenges in GRB luminosity variability, ultimately extending the cosmic distance ladder in a robust way. By addressing redshift evolution and reducing systematic uncertainties, GRBs can serve as reliable high-redshift distance indicators, offering critical insights into current cosmological tensions.
\end{abstract}

\maketitle

\section{Introduction}\label{sec:intro}

The $\Lambda$CDM concordance model has provided several decades of successful predictions both at the astrophysical and cosmological regimes of physics \cite{Peebles:2002gy,Copeland:2006wr}. However, the framework of cold dark matter \cite{Baudis:2016qwx,XENON:2018voc}, together with a cosmological constant realization of dark energy \cite{Riess:1998cb,Perlmutter:1998np} and a general relativity perspective of gravitation \cite{Misner:1973prb,Weinberg:1972kfs} has increasingly been challenged by new precision measurements. There have been theoretical and observational issues with CDM since its inception, ranging from the fundamental physics source of the cosmological constant \cite{Weinberg:1988cp} to the more general question of its ultra-violet completeness \cite{Addazi:2021xuf} and the direct observations of CDM candidates \cite{LUX:2016ggv,Gaitskell:2004gd}, besides other challenges to the concordance model \cite{DiValentino:2025sru}. Over the last decade, the precision measurements of cosmological parameters in the local Universe \cite{Riess:2021jrx,Blakeslee:2021rqi} have seemingly turned out to be dissimilar to those based on early-time observational data. An important caveat in this description is that early measurements of cosmological parameters are inherently based on the $\Lambda$CDM model and can only be disentangled through competing models \cite{ACT:2023kun,Schoneberg:2022ggi}.

This contrast is most poignantly observed by the local measurements of SNe Ia in the Pantheon+ sample, which, when calibrated with Cepheid hosts for absolute distances, report a Hubble expansion rate of \cite{Riess:2021jrx} which is strengthened by measurements using the surface brightness fluctuation method of early-type galaxies \cite{Blakeslee:2021rqi}.
These measurements are direct in nature in that they do not require a model of cosmology such as $\Lambda$CDM, which is in contrast to measurements centered on the early Universe \cite{Planck:2018vyg}. In the early Universe regime, one of the latest releases of analysis of the cosmic microwave background radiation by the Atacama Cosmology Telescope (DR6) gives the Hubble constant, $H_0^{\rm ACT} = 68.3\pm 1.1 {\rm kms}^{-1}{\rm Mpc}^{-1}$ \cite{ACT:2023kun}. In contrast, predictions of  using the combination of Big Bang nucleosynthesis data with baryonic acoustic oscillation (BAO) measurements give values shown in \cite{Schoneberg:2022ggi}. The statistical distance between these types of measurements has prompted a reexamination of the data analysis methods used in these studies and the underlying physical model being employed.

To confront these issues, there has been a growing chorus of independent distance indicators and expansion rate measurement techniques. An important technique that is quickly increasing in precision is that based on gamma ray bursts (GRB) measurements \cite{2009MNRAS.400..775C,2010MNRAS.408.1181C,Dainotti:2013cta,postnikov2014,Dainotti:2023pwk,Dainotti2023alternative}. GRBs offer the possibility of reaching far higher redshifts \cite{Banados:2017unc,2011ApJ...743..154C} as compared with more traditional techniques with possible detections up to and over $z \gtrsim 20$ \cite{Lamb:2000qe,Bloom:2008ua}. Here, GRBs offer events with extreme brightness intensities as well as redshift coverage that overlaps quite well with other data sets, including SNe Ia and BAO samples, which makes the issue of calibration more critical. However, to standardize GRBs as cosmological standard candles, it is necessary to use them as cosmological standard candles \cite{Dainotti2008,2010ApJ...722L.215D,Dainotti:2011ue,Dainotti:2013cta,Dainotti2015, Dainotti2016, 2017A&A...600A..98D,2020ApJ...905L..26D,Dainotti:2022jzv,Cao:2022wlg} which occurs after a selection process happens among the GRB classes and their morphological properties. Traditionally, the calibration process has been based on a two-dimensional (2D) relation \cite{2008MNRAS.391L..79D,Dainotti:2013fra} between the luminosity at the end of the plateau emission and its rest frame duration. 
This relation can be ascribed to processes occurring due to the magnetar emission \cite{Rowlinson:2014dja,2015JHEAp...7...73D,Rea:2015gna,Stratta:2018xza,DallOsso:2023gdk}. 
In order to reduce the scatter of this relation, a third parameter has been added: the peak prompt luminosity thus creating the proposed three-dimensional (3D) Dainotti relation \cite{2016ApJ...825L..20D,2017ApJ...848...88D,Dainotti:2020azn}.  Moreover, the 3D relation is featured as a natural tighter relation compared to the 2D one. Its scatter decreases when given morphological properties of the plateau emission are considered.

The redshift overlap between GRBs and other late time measurements means that the calibration of GRBs can be verified by comparing the distance luminosity obtained with these relationships. Another way to utilize this overlap is to adopt the independent expansion profile such as the Hubble profile, as the calibrated profile and then reconstruct the calibration relation parameters using this calibrated Hubble profile. However, the exact redshift points are naturally distinct from the calibrated points, and so the Hubble diagram and its associated uncertainties must be interpolated. Moreover, this profile needs to be mapped to its $D_L(z)$ to complete the calibration profile. There is a variety of ways in which these redshift points can be interpolated including the use of a fiducial model. However, to remove to the fullest extent possible the role of systematics, we use model-independent approaches whereby the role of physical models is transferred to statistical inference and learning processes. We would like to clarify that our model-independent approach is independent of a cosmological model, but not of a statistical model. Indeed, the use of a neural network trained on the Pantheon+ dataset, the $\chi^2$ loss function, and the Gaussian likelihood used in the MCMC analysis all have the same embedded assumptions, particularly regarding the statistical properties of the data, such as Gaussianity of the errors. This is a starting point of the work, because we are aware of the non-Gaussianity of the Pantheon+ data as shown in \cite{Dainotti2024JHEAp..41...30D} that the residual of the observed distance moduli and the theoretical one are not Gaussian in the Pantheon+ data. Work has already been done using Gaussian processes (GP) \cite{2012JCAP...06..036S,Shafieloo:2012ht,Seikel:2013fda, Yennapureddy:2017vvb, Gomez-Valent:2018hwc, Li:2019nux, Liao:2019qoc, Keeley:2020aym, Renzi:2020fnx,OColgain:2021pyh} whereby points in the calibration data set are assumed to be Gaussian distributions and are used to train non-physical hyperparameters in a covariance, or kernel, relation which can then be employed to estimate distinct points in the redshift interval. This approach was employed in \cite{Favale:2024lgp} where the systematics of the calibration relation were investigated in a model-independent way. In this way, constraints on the 2D and 3D calibration relation parameter sets can be obtained using a Markov chain Monte Carlo (MCMC). The results of this work are interesting in that the constraints obtained are relatively narrow, namely $\Delta_{\sigma_v}=0.03/0.21=14\%$. In addition, there is a difference of 1.5 in relation to the z-score for the $\sigma_v$.

On the other hand, GP has a series of issues such as the selection of the kernel function as well as overfitting tendencies for low redshifts which is known to restrict models and parameters \cite{Hwang:2022hla,LeviSaid:2021yat,Keeley:2020aym,Briffa:2020qli,Bernardo:2021qhu,Benisty:2020kdt,Perenon:2021uom,Bernardo:2021cxi} despite efforts to overcome these deficiencies \cite{Bernardo:2021mfs,Bernardo:2022pyz}. Another approach which has a lower dependence on pipeline architecture is that of artificial neural networks (ANN) where neurons are modeled on their biological analog, and are organized into layers through which data points are transformed to signals that pass through the network. Thus, the calibration data set would take the form of a training data set by which the ANN would be optimized to accept redshifts and produce Hubble parameter points and their associated uncertainties \cite{aggarwal2018neural,Wang:2020sxl,Gomez-Vargas:2021zyl}. This technique has the advantage that it provides one of the most competitive model-independent approaches to reconstructing Hubble profiles without any meaningful reliance on statistical or model choices. One early example of this method is in \cite{Wang:2019vxv} where the Hubble diagram was reconstructed using simulated and cosmic chronometer data. The approach was expanded in \cite{Dialektopoulos:2021wde} to include BAO data and was also applied to large scale structure data. In \cite{Dialektopoulos:2023dhb}, the ANN pipeline was further nuanced with the full covariance for SNe Ia samples for the training phase of the ANN pipeline. Recently, ANNs were used to construct functional derivatives \cite{Mukherjee:2022yyq} which has opened the way for the reconstruction of cosmological models \cite{Dialektopoulos:2023jam}.

The question of systematics or the over-reliance on a particular model plagues the analysis of many types of observational pipelines. Model independent approaches, and ANNs in particular, offer novel approaches to tackling this potential question. An instance of such studies involves the probing of potential evolution or a variation over the concordance model of the absolute magnitude of SNe Ia events. The value of the absolute magnitude of SNe Ia events is a constant in the concordance model but potential variations have been widely modeled \cite{Mukherjee:2024akt,Zuckerman:2021kgm,Camarena:2019moy,Camarena:2021jlr,Marra:2021fvf,Alestas:2020zol,Alestas:2021xes,Alestas:2021luu,Dainotti:2021pqg}. However, it has been argued that the statistical significance of deviations was relatively moderate. On the other hand, this is an excellent test bed for using model-independent approaches to probe systematics. The use of model-independent methods to probe the possible evolution of the absolute magnitude has been wide and extensive \cite{Camarena:2019rmj,Mukherjee:2021kcu,Mukherjee:2020ytg,Dinda:2022jih,Gomez-Valent:2021hda,Favale:2023lnp,Banerjee:2023evd}, while there are less works on using ANNs to probe this question \cite{Shah:2024slr,Benisty:2022psx,Mukherjee:2024akt}. In \cite{Shah:2024slr}, a new toolbox called LADDER is introduced to reconstruct cosmological parameters using training data of a similar type, while in \cite{Benisty:2022psx} BAO data was used to establish the comoving sound horizon at drag epoch, which is then used to invert the SNIa calibration relation in order to use event data to model the absolute magnitude. This work reaffirmed the significance of variations in the current data set but the analysis provided a way in which to explore such potential systematics. Finally, \cite{Mukherjee:2024akt} uses the cosmic chronometer Hubble data to build a baseline on which SNe Ia data can be used to probe the absolute magnitude at different redshifts.

In this work, we extend the approach established using SNe Ia measurements to GRB data sets to constrain the calibration relation parameters using ANNs. Using luminosity distance measurements to directly establish a baseline Hubble diagram, the 2D and 3D GRB calibration relations can be used in conjunction with an MCMC analysis approach to resolve the values of the two sets of calibration parameters in a cosmological model-independent way. This will remove the dependence on fiducial models and a large portion of systematics with it. In that spirit, BAO data is omitted from the study to circumvent the known model dependence issues of these data sets. As for the GRB data we use the Platinum compilation \cite{Dainotti:2020azn} which consists of 50 long GRBs in the redshift interval $0.553 \leq z \leq 5.0$ which has been built so that the plateau emission properties have been well-defined. This sample has been used to further reduce the scatter of the 3D Dainotti relation and it has been presented the first time in \cite{Dainotti:2020azn}. We use this sample the compare the 2D and 3D calibration relations in a model independent way in the context of their calibration parameters. 

The work is organized as follows: In Sec.~\ref{sec:ANN} we introduce out implementation of ANN-based neural networks. In Sec.~\ref{sec:data} we discuss the GRB data sample under consideration here, while in Sec.~\ref{sec:2d_3d_calib} the 2D and 3D Dainotti relations are introduced. The ANN results are shown in Sec.~\ref{sec:ANN_output_2d_3d} for different combinations of initial priors. A reduction of the scatter on the parameters is observed and it is then confronted in Sec.~\ref{sec:redshift_evo}. The work is compared with an analogue analysis that was performed using GP in Sec.~\ref{sec:comparison_discussion}, followed by the cosmological inference based on the calibrated GRBs in Sec.\ref{sec:cosmo}. Finally, the implications of the entire work are discussed in the conclusion in Sec.~\ref{sec:conclusion}.

\section{Artificial Neural Network Framework} \label{sec:ANN}

The implementation of the ANN method is inspired by its biological analogue wherein neurons are constructed into a network and organized into layers \cite{2015arXiv151107289C}. An input layer through which the sample data enters the network, while the sample calibration parameters are produced by an output layer. Through a series of consecutive layers that contain neurons, called hidden layers, internal hyperparameters, which are nonphysical, are optimized through a learning process in which real data is used to establish an interpolated Hubble diagram \cite{Dialektopoulos:2021wde}. In this way, a cosmology-free baseline can be used to constrain the 2D and 3D GRB calibration parameters.

For our case, the input layer will accept redshift values while the output layer will produce the luminosity distance for that redshift and its associated uncertainties. Through this architecture, signals will traverse the entire network which can be used to train the pipeline hyperparameters, thus optimising the ANN outputs through training data. As an illustrative example, Fig.~\ref{fig:ANN_structure} shows a two-layer architecture of neurons denoted by $\mathfrak{n}_k$ and $\mathfrak{m}_k$, which accepts redshift data points and outputs a generic cosmological parameter $\Upsilon(z)$ together with its associated uncertainty $\sigma_\Upsilon^{}(z)$. This ANN pipeline is structured in such a way that a linear transformation (composed of linear weights and biases) is applied for each of the different layers that organize the neurons.

	\tikzset{%
		every neuron/.style={
			circle,
			fill=green!60,
			minimum size=32pt, inner sep=0pt
		},
		mid neuron/.style={
			circle,
			fill=blue!30,
			minimum size=32pt, inner sep=0pt
		},
		last neuron/.style={
			rectangle,
			fill=red!30,
			minimum size=32pt, inner sep=1pt
		},
		neuron missing/.style={
			draw=none,
			fill=none,
			scale=4,
			text height=0.333cm,
			execute at begin node=\color{black}$\vdots$
		},
		circle node/.style={
			circle,
			fill=orange!30,
			minimum size=32pt,
			inner sep=2pt,
		}
	}
	
	\begin{figure}
		\centering
		\begin{tikzpicture}[shorten >=1pt,->,draw=black!50, node distance=\layersep]
		\tikzstyle{annot} = [text width=5em, text centered]
		
		\foreach \m/\l [count=\y] in {1}
		\node [every neuron/.try, neuron \m/.try] (input-\m) at (0,-1.5*\y) {};
		
		\foreach \m [count=\y] in {1,2,missing,3}
		\node [mid neuron/.try, neuron \m/.try ] (hidden1-\m) at (2.5,2-\y*1.5) {};
		
		\foreach \m [count=\y] in {1,2,missing,3}
		\node [mid neuron/.try, neuron \m/.try ] (hidden2-\m) at (5.2,2-\y*1.5) {};
		
		\foreach \m [count=\y] in {1}
		\node [last neuron/.try, node distance=1cm] (output-\m) at (7.52,-1.5) {};  
		
		\foreach \name / \y in {1}
		\path[yshift=-.1cm] node[above] (input+\name) at (0,-1.6\name) {\large$z$};
		
		\foreach \l [count=\i] in {1,2,k}
		\node at (hidden1-\i) {\large$\mathfrak{n}_\l$};
		
		\foreach \l [count=\i] in {1,2,k}
		\node at (hidden2-\i) {\large$\mathfrak{m}_\l$};
		
		\node [last neuron/.try, node distance=1cm] (output-node) at (8.52,-1.5) {\large $P(\Upsilon) \sim \mathcal{N}(\mu, \sigma)$};
		
		\node [circle node/.try] (mu) at (11,-0.5) {\large $\Upsilon(z)$};
		\node [circle node/.try] (sigma) at (11,-2.5) {\large $\sigma_\Upsilon(z)$};
		
		\foreach \i in {1}
		\foreach \j in {1,2,...,3}
		\draw [->] (input-\i) -- (hidden1-\j);
		
		\foreach \i in {1,2,...,3}
		\foreach \j in {1,2,...,3}
		\draw [->] (hidden1-\i) -- (hidden2-\j);
		
		\foreach \i in {1,2,...,3}
		\foreach \j in {1}
		\draw [->] (hidden2-\i) -- (output-\j);
		
		\draw [->] (output-node) -- (mu);
		\draw [->] (output-node) -- (sigma);
		
		\draw[thick, color=black, domain=7.25:9.75, samples=100, -] 
		plot (\x,{1.2*exp(-((\x-8.5)^2)/0.5)}) ;
		
		\draw[->] (7,0) -- (10,0) node[right] {\large $x$};
		
		\draw[->] (8.5,-0.1) -- (8.5,1.55) node[above] {\normalsize $P(\Upsilon)$};
		
		\draw[dashed, color=red, -, line width=0.1em] (8.5,0) -- (8.5,1.2);
		\draw[dotted, color=blue, -, line width=0.1em] (8.5,0.79) -- (8.95,0.79);
		\draw[dotted, color=blue, -, line width=0.1em] (8.96,0.) -- (8.95,0.79);
		
		\node at (8.25,0.5) {\color{red} \large $\mu$};
		\node at (8.75,-0.2) {\color{blue} \large $\sigma$};
		
		\node (1) at (0,3){\large Input};
		\node (1) at (0,2.55){\large layer};
		
		\node (1) at (5,3){\large Hidden};
		\node (1) at (5,2.55){\large layers};
		
		\node (1) at (10,3){\large Output};
		\node (1) at (10,2.55){\large layer};
		
		\end{tikzpicture}
		\caption{A two-layer ANN architecture is shown, where the input is the redshift of a cosmological parameter $\Upsilon(z)$, and the output is the probability distribution $P(\Upsilon) \sim \mathcal{N}(\mu = \Upsilon(z), \sigma = \sigma_\Upsilon(z))$. The parameters $\mu$ and $\Sigma$ are derived from the output distribution.}
		\label{fig:ANN_structure}
	\end{figure}
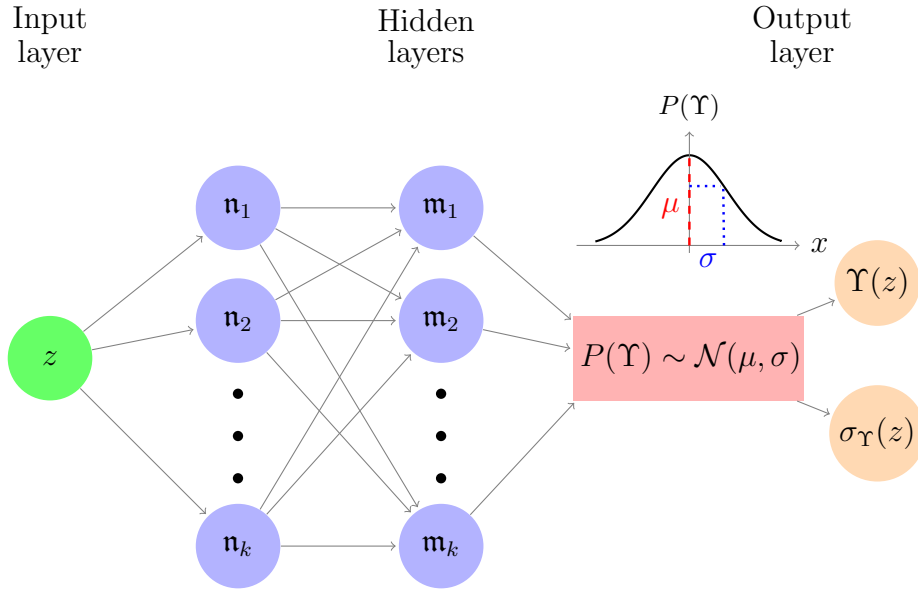

The complex relationships that feature the data can be modeled in the organisation of the system of neurons through the use of a triggering activation function which has an impact over large numbers of neurons. In our study, we use the Exponential Linear Unit (ELU) \cite{2015arXiv151107289C} activation function, defined by
\begin{equation}
    f(x) = 
    \begin{cases} 
           {x} & \text{if } x>0 \\
        {\alpha(e^x-1)} & \text{if } x \leq 0
    \end{cases}\,,
\end{equation}
where $\alpha$ is a positive hyperparameter that adjusts the value at which the ELU saturates for negative inputs, and which assumes unity for other instances.

The activation function does not act on positive inputs and tends to map negative ones closer to unity for progressively negative values. The choice in activation function defines the network type, such as multilayer perceptron, convolutional, recurrent, etc, while the equivalent choice for the output layer is normally related to the problem domain, such as classification or regression, more details can be found in \cite{DeepLearning-book}.

Together, the linear transformations and activation functions produce a large number of hyperparameters which must be optimized through a training process. These parameters are not physical but together they produce a pipeline that can closely mimic physical processes. There are even more hyperparameters that emerge due to the relationships between the neurons themselves, which gives the ANN system a wide breadth of freedom. The training process optimizes the hyperparameter values through successive iterations of a training set which sets the baseline of the system. The goal of the process is for the predicted results $\hat{\Upsilon}$ to resemble the training data $\Upsilon$, as closely as possible. This minimization procedure is characterized through a loss function which defines the difference between these two sets. The function is minimized through regular fitting procedures such as gradient descent. For our analysis, we adopt Adam's algorithm \cite{2014arXiv1412.6980K} which is a slightly more efficient variant of the gradient descent method.

The straightforward difference between the predicted ($\hat{\Upsilon}$) and training ($\Upsilon$) output data summed for every redshift is called the L1 loss function. This is inspired by the log-likelihood function in MCMC for uncorrelated data, and is a very popular choice of the ANN architecture. While other choices exist such as mean square error (MSE) and smoothed L1 (SL1), for cosmological data L1 appears to be the most efficient from previous studies \cite{Dialektopoulos:2021wde, Mukherjee:2022yyq}. It is through the loss function that the complexity within the data is embodied in the ANN architecture. In our case, we are inspired again by the log-likelihood of MCMC analyses where a covariance matrix $C$ is incorporated through the $\chi^2 $ loss function \cite{Mukherjee:2024akt, Dialektopoulos:2023dhb}
\begin{equation}
    {\rm L_{\chi^2}} = \sum_{i,j} \left[\Upsilon_{\rm obs}(z_i) - \hat\Upsilon_{\rm pred}(z_i)\right]^\text{T} \mathcal{C}_{ij}^{-1} \left[\Upsilon_{\rm obs}(z_j) - \hat\Upsilon_{\rm pred}(z_j)\right] \,, \label{eq:chi2_loss}
\end{equation}
where $\mathcal{C}_{ij}$ are the components of the covariance matrix that includes statistical noise and systematics.

\begin{figure}
    \centering
    \includegraphics[width = 0.495\textwidth]{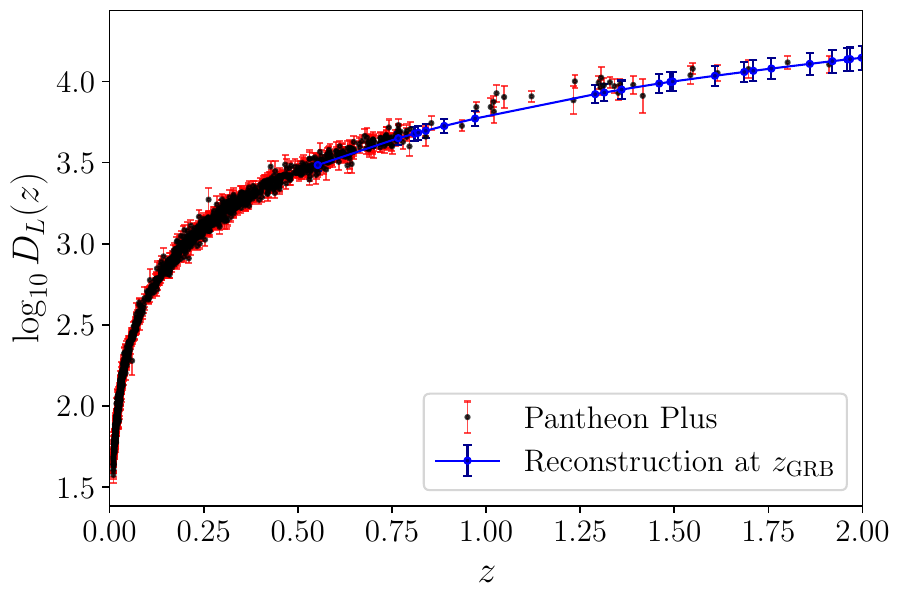}
    \includegraphics[width = 0.495\textwidth]{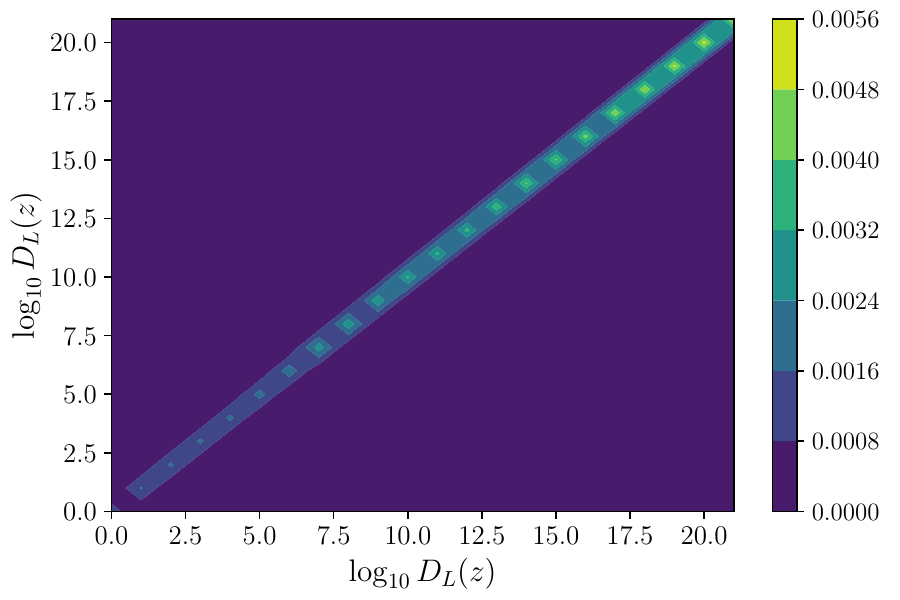}
\caption{ANN reconstruction of the Pantheon+ SNIa logarithmic  $D_L(z)$, $\log_{10} D_L(z)$, as a function of the redshift $(z)$ in the left panel. The right panel shows the covariance matrix between ANN reconstruction of the Pantheon+ SN-Ia $\log_{10} D_L(z)$. The color bar on the right shows the covariance  $\text{cov}[\log_{10} D_L(z_i), \, \log_{10} D_L(z_j)] $ at redshifts $z_i$ and $z_j$ respectively.}
    \label{fig:dl_rec}
\end{figure}

ANNs can approximate any continuous function for a finite number of neurons and one hidden layer, provided the activation function is continuous and differentiable \cite{HORNIK1990551}, meaning that they are suitable for the setting of cosmological data. In our study, we utilize \texttt{PyTorch} \footnote{\url{https://pytorch.org/docs/master/index.html}} for this implementation to reconstruct $D_L(z)$ from the Pantheon+ SNe-Ia compilation. The value of the Pantheon+ apparent magnitude is set to $M_B=-19.35$, following \cite{Mukherjee:2024akt}. The optimal configuration features two hidden layers with 128 neurons each. To optimize the network’s performance, we split the dataset into training (70\%)
and validation (30\%) sets. We incorporate the covariance matrix of the Pantheon+ dataset into the training algorithm, where we opt to learn the noise information simultaneously with the observational data via minimization of the $\chi^2$ loss, defined in Eq. (\ref{eq:chi2_loss}).  To speed up the pipeline convergence rate we use GPUs and batch normalization \cite{2015arXiv150203167I} prior to every layer. We consider training batch sizes of 32, and the covariance sub-matrices equivalent for every training batch are carefully selected ensuring that these sub-matrices remain positive semi-definite (as established in \cite{Mukherjee:2024akt}). We plot the reconstructed mean and 1$\sigma$ uncertainties of $\log_{10}D_L$(z) in the left panel of Fig. \ref{fig:dl_rec}. The covariance between the reconstructed function values at different GRB redshifts is shown in the right panel of Fig. \ref{fig:dl_rec}.

\begin{figure}[h!]
    \centering
    \includegraphics[width=0.45\textwidth]{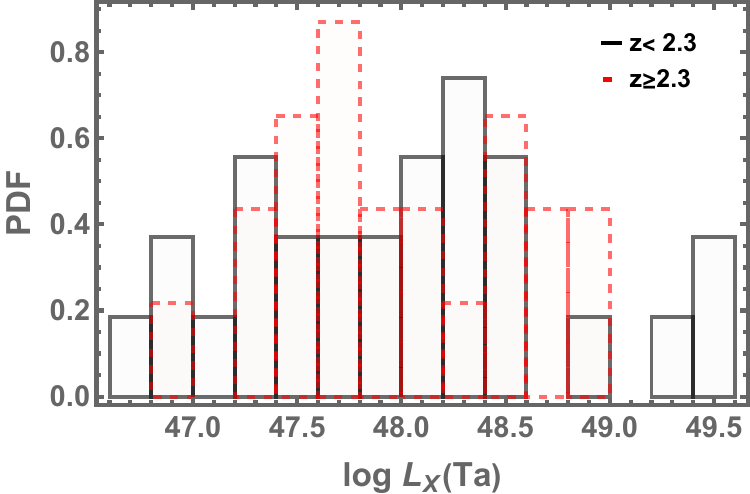}
    \caption{Differential distributions of the $L_X(T_a)$ variable of the samples at low-z ($z<2.3$) shown in continuous line and with $z \geq 2.3$ with red dashed line.}
    \label{fig:dist_compare}
\end{figure}

\section{Data sample}  \label{sec:data}

We have taken the sample from \cite{Dainotti:2020azn}, where they have analyzed all GRBs presenting X-ray plateau afterglows detected by {\it Swift} from January 2005 up to August 2019 using the BAT + XRT light-curves gathered from the Swift web page repository \cite{2009GCN..9724....1M,2010A&A...519A.102E}. These GRBs have known redshifts spectroscopic or photometric, available in \cite{Xiao:2009dr} the Greiner web page \footnote{http://www.mpe.mpg.de/jcg/grbgen.html}, and in the Gamma-ray Coordinates Network (GCN) circulars and notices \footnote{http://gcn.gsfc.nasa.gov/}. Only GRBs with precise redshift measurements have been considered. We point out that the long GRB sample has been built from the whole sample, subtracting the short GRBs with extended emission, the X-ray flashes, which have unusually soft spectra and greater fluence in the X-ray band (2-30 keV) than in the $\gamma$-ray band (30-400 keV), and ultra-long GRBs (ULGRBs) with a very long prompt duration ($>1000$ s), the GRBs associated with SNe Ib/c, which means that a GRB belonging to the long class cannot be a part of the other classes here mentioned. To further reduce the intrinsic scatter of the fundamental plane and increase its reliability as a cosmological probe, we define a subsample of the Gold Sample (defined in \cite{Dainotti:2016iqn}), the Platinum Sample. This includes well-sampled light-curves of 50 GRBs, which present the following features: 
\begin{enumerate}[left=0pt]
    \item[(a)] GRBs should have a relatively flat plateau, whose angle between the end of the plateau and the beginning should not be larger than $41^\circ$;  
    \item[(b)] $T_{X}$ (end time of the plateau emission) should not be inside a large gap of the data, thus having large uncertainty;
    \item[(c)] a small plateau duration ($<500$ s) and with gaps after it should be discarded. This could mean that the plateau phase is longer than the one observed. This latter condition ensures that we have no ambiguity on the existence of the plateau itself, since where a plateau is masked out from the prompt emission is unclear if a plateau exists; and
    \item[(d)] cases of GRBs with flares and bumps at the start and during the plateau phase should also be eliminated.
\end{enumerate}

\section{The 2D and 3D Dainotti relations} \label{sec:2d_3d_calib}

We take into account GRBs which can be described by the phenomenological model \cite{2008AIPC.1000...24W}.  We follow the criteria for the GRB sample selection considered in \cite{Srinivasaragavan:2020isz} and \cite{Dainotti:2020azn}, and we use the platinum sample detailed in \cite{Dainotti:2020jkj}. We fit this sample with the Willingale functional form for ${f(t)}$, which reads
\begin{equation}
    f(t) = \begin{cases}
    F_i \exp{\left( \alpha_i \left( 1 - \frac{t}{T_i} \right) \right)} \exp{\left(-\frac{t_i}{t} \right)} & {\rm for} \ \ t < T_i \\
    F_i \left(\frac{t}{T_i} \right)^{-\alpha_i}\exp{\left( -\frac{t_i}{t} \right)} & {\rm for} \ \ t \ge T_i \, ,\newline
    \end{cases}
    \label{eq: fc}
\end{equation}
modelled for both the prompt (index `i=\textit{p}') ${\gamma}$\,-\,ray and initial hard X-ray decay and for the afterglow (`i=\textit{X}'), so that the complete lightcurve ${f_{tot}(t) = f_p(t) + f_X(t)}$ contains two sets of four free parameters ${(T_{i},F_{i},\alpha_i,t_i)}$, where ${\alpha_{i}}$ is the temporal power-law decay index and $T_i$ is the end time of the prompt and the plateau emission, respectively, while the time ${t_{i}}$ is the initial rise timescale. The transition from the exponential to power-law occurs at the point ${(T_{i},~ F_{i}e^{-t_i/T_i})}$, where the two functions have the same value and this point marks the beginning of the plateau. Most cases are fitted by fixing the rise time.
Using these criteria, we fit 222 light-curves and from these 222 we obtain the subsample of 50 GRBs. The peak prompt luminosity at 1 second, $L_{\rm peak}$, and the X-ray luminosity measured in the final part of the plateau phase, $L_X$, have been calculated from
\begin{equation}
    L= 4 \pi D_L^2(z) \, F (E_{\rm min}, \, E_{\rm max}, \, T_{X}) \cdot K  \,  ,
    \label{Lpeak}
\end{equation}
where $K$ is the $K$-correction for the cosmic expansion \cite{2001AJ....121.2879B} and $F (E_{\rm min},E_{\rm max},T^{*}_{X})$ is the energy flux, $T_{X}$ is the time measured in the observer frame at the end of the plateau and $E_{\rm min}$, $E_{\rm max}$ are the minimum and maximum energies of the band pass of the given satellite instrument in question.
To calculate $L_{\rm peak}$ one substitutes the flux $F$ with $F_{\rm peak}$, which is the $\gamma$-ray flux in $1$ s interval ($\mathrm{erg}$ $\mathrm{cm}^{-2} \mathrm{s}^{-1}$) measured at the peak of the prompt emission, while to calculate $L_{\rm X}$, one uses the flux $F_{X}$, measured in X-rays at the end of the plateau. $D_L(z)$ is written as
\begin{equation} \label{flatdistanceluminosity}
    D_L(z)=(1+z)\frac{c}{H_{0}}\int_{0}^{z} \frac{dz'}{\sqrt{\Omega_{m}(1+z')^3+(1-\Omega_{m}})} \, ,
\end{equation}
and it is computed for a given redshift in the flat $\Lambda$CDM cosmological model, according to which we have an energy equation of state $w=-1$, the dark matter density, $\Omega_m=0.3$, and $H_0=70\, {\rm km s}^{-1} {\rm Mpc}^{-1}$ and where $c$ is the speed of light. For GRBs whose spectrum is fitted by a simple power-law the K-correction is given by $K=(1+z)^{(\beta - 1)}$, where $\beta$ is the spectral index of the plateau in the X-ray band \cite{2009GCN..9724....1M,2010A&A...519A.102E}. With these two equations and the fitting of the time at the end of the plateau emission, we can build the 2D and 3D relations as
\begin{equation}
    \log L_X = C_0 + a\cdot \log T^{*}_{X} + b \cdot( \log L_{\rm peak}), \label{eq:isotropic}
\end{equation} 
where $C_0$  is the normalization constant and denotes energy, the $a$ parameter implies the relation between the luminosity at the end of the plateau, $L_X$, and the correspondent rest frame time $T_X$ and $L_{\rm peak}$ is the prompt peak luminosity. \cite{2008MNRAS.391L..79D}  discovered a correlation for long GRBs between  $L_X$, and $T^{*}_X$ (named the two-dimensional Dainotti relation in X-rays or $L_X-T^{*}_X$ relation), where the rest frame quantities are denoted with $\star$ symbol. The established behavior is $\log L_X = \log a + b \log T^{*}_X$, where  $C_0$ (the normalization) and $a$ (slope, roughly $-1$) are obtained from fitting the observations. The more luminous the plateau phase, the shorter its duration, i.e., the quicker it consumes its energy. The $L_X-T^{*}_X$ correlation was updated with 77 GRBs \cite{2010ApJ...722L.215D}. This correlation, being related to the plateau phase, whose mechanism is still unknown, has been useful in testing theoretical models \cite{Gehrels:2009cr,Cannizzo:2010jx,uehara2010infrared,2014MNRAS.445.2414V,2014MNRAS.442.3495V,Rowlinson2014,Hascoet:2013bma,Rea:2015gna,Stratta:2018xza}. Moreover, analyzing other prompt and afterglow parameters led to new correlations between $L_X$ and the mean prompt luminosities, $L_{\gamma}(T)=E_{\rm iso}/T$, ($E_{\rm iso}$ is the isotropic radiated energy in the prompt) computed at different times, $T$ \cite{Dainotti2011}. Refs \cite{Dainotti:2011ue,Dainotti:2013fra} investigated further the relation and corrected for selection biases and redshift evolution. \cite{Dainotti2015} showed that the $L_{\rm peak}-L_X$ relation has a higher Spearman correlation coefficient, $\rho=0.79$, than the one, $\rho=0.60$, obtained for the averaged prompt luminosities, $L_{\rm prompt}-L_X$ correlation, for the same sample. Since the prompt and afterglow relations have in common the plateau luminosity, \cite{Dainotti2016} combined these two relations, $L_{\rm peak}-L_X$ and $L_X- T^{*}_X$, finding a tighter relation that increases the utility of GRBs as standard candles. This reduced the scatter (up to 54\% if compared with the two-dimensional correlations) within long GRBs with relatively flat plateaus, called the gold sample. These results were verified using Monte Carlo simulations to prove that the 3D correlation was not a random effect due to the sample used. The intrinsic nature of this relation was proven in \cite{dainotti2017b} and updated in \cite{Dainotti:2020azn}. According to the magnetar model, the value of the slope needs to be around $-1$ \cite{Rowlinson:2014dja,Stratta:2018xza}. Thus, the prior used in the subsequent analysis that $a<0$ is well-motivated. Regarding the $b$ parameter, this denotes how much kinetic energy is transferred from the prompt to the afterglow emission. Since the more luminous the prompt, the more kinetic power is transferred, the $b$ parameter must be $b>0$. This justifies the priors on this parameter indeed as well. On the other hand, after 21 years of observations of the prompt-afterglow parameters, we did not find any indication for a negative correlation, thus we find reasonable to consider this realistic parameter space.

Given that our scope is to extrapolate $D_L(z)$ at high-z, beyond the redshift of SNe Ia ($z=2.26$), we need to guarantee that the distribution of the $L_X(T_a)$ derived from Eq. \ref{eq:isotropic} and from which $D_L(z)$ at low z ($z < 2.3$) is obtained from the relation via Eq. \ref{flatdistanceluminosity}, should be drawn by the same parent population of $L_X(T_a)$ at higher redshifts ($z \geq 2.3$). Thus, we have checked the distributions of the $L_X(Ta)$ of the platinum sample at $z \leq 2.3$, shown in a continuous black line, is similar to the distribution at $z \geq 2.3$ shown in dashed red, see Fig. \ref{fig:dist_compare}. Indeed, the inferred Kolmogorov Smirnov Test is of 0.92, thus showing that the two distributions are statistically drawn by the same population.

\section{Calibration of the 3D and the 2D  Dainotti relations} \label{sec:ANN_output_2d_3d}

The calibration of the parameters laid out in the 3D Dainotti relation in Eq.~\eqref{eq:isotropic} are explored in this section. We take a similar approach to \cite{Favale:2024lgp} of first establishing a data-driven reconstruction of the luminosity distance, albeit using our trained ANN architecture, discussed in Sec. \ref{sec:ANN} on the Pantheon+ SN-Ia data. This network model is then used to infer the theoretical $D_L(z)$ at GRB sample redshifts $z_{\rm GRB}$, so that the parameters of the GRB calibration relations can be constrained using traditional MCMC methods. 

The log-likelihood to be evaluated at each step of the MCMC analysis is
\begin{equation} \label{eq:lnlike}
    \ln \mathcal{L} = -\frac{1}{2} \left[ \sum_{i=1}^N \ln \left\lbrace 2\pi \left( \sigma^2_{{\rm rec}, i} + \sigma^2_{\rm int}\right) \right\rbrace + \chi^2_{\rm GRB}\right] \,
\end{equation} 
where,
\begin{equation}
    \chi^2_{\rm GRB} = \sum_{i=1}^N \frac{\left[ \log_{10} D_L^{\rm rec}(z_i) - \log_{10} D_L^{\rm th}(z_i, a, b, C_0) \right]^2}{\sigma^2_{{\rm rec}, i} + \sigma^2_{\rm int}} \, ,
\end{equation}
such that
\begin{equation} \label{eq:dl_theo}
    \log D_L^{\rm th} = a_1 \log T_X^{*} + b_1 (\log F_{\rm peak} + \log K_{\rm prompt})) + c_1 + d_1 (\log F_X + \log K_{\rm plateau}) \, 
\end{equation} 
denotes the theoretical value for the $\log D_L(z)$, in units of cm. Here, $a_1 = a/2(1 - b)$, $b_1 = b/2(1 - b)$, $c_1 = ((b - 1) \log(4\pi) + C_0 )/(2(1 - b))$ and $d_1 = -(1 - b)/2$ respectively. This derivation is taken from \cite{Dainotti:2023pwk}. Note that, $\log_{10} D_L^{\rm rec}$ and $\sigma^2_{{\rm rec}, i}$ are the ANN reconstructed mean and 1$\sigma$ values of the $\log D_L(z)$(shown in Fig. \ref{fig:dl_rec}) from Pantheon+ SN-Ia data at the $N$ GRB observations. The corresponding covariance matrix for reconstructed logarithmic luminosity distances is obtained as follows,
\begin{equation}
    C_{{\rm rec}, ij} = \frac{1}{n} \sum_{k=1}^{n} (x_{k, i} - \overline{x}_i)(x_{k, j} - \overline{x}_j) 
\end{equation} 
where $x_{k,i}$ is the reconstructed values of $\log D_L^{\rm rec}(z)$ at the $i$-th redshift for $k = 1, ..., n$ realizations obtained from the Pantheon+ trained ANN pipeline.

In this way, the constraints on the 3D Dainotti relation will be achieved in a cosmological model-independent manner. Through this way we have overcome the so-called circularity problem. This is achieved using the luminosity distance relations derivable from Eq. \eqref{eq:dl_theo} to bridge these two types of data sets together. This will build on the constraints achieved in Ref.~\cite{Favale:2024lgp}, where a similar method was adopted except for using a GP pipeline on the cosmic chronometer Hubble data to produce the Hubble diagram. We aim to produce an independent examination of the results obtained by substituting an ANN for this component of the analysis. The results will be compared in section \ref{sec:comparison_discussion}.

In our analysis, we make use of the Platinum sample of GRB observational measurements and Pantheon+ trained ANN pipeline to maximize the log-likelihood, defined in \eqref{eq:lnlike}, using a traditional MCMC sampler, \texttt{emcee}, for the correlated $N_{\rm GRB}$ data points. In what follows, we undertake a two-fold method where we keep the five parameters defining the fundamental plane (i.e. $\left\lbrace \log T_X^{\star}, \log F_X, \log K_{\rm plateau}, \log F_{\rm peak}, \log K_{\rm prompt} \right\rbrace$) as (a) fixed; (b) free to vary as nuisance parameters in the MCMC analysis. To verify the robustness of this approach within the ANN framework, we consider different prior choices on the calibration parameters $a$, $b$, $C_0$, and $\sigma_{\rm int}$. We perform this verification, taking both Gaussian and flat priors with means fixed on fits from the fundamental plane, with gradually increasing the standard deviation in multiples of $\sigma$ for both scenarios.

\squeezetable
\begin{table}[t]
{\renewcommand{\arraystretch}{1.5} \setlength{\tabcolsep}{12 pt} \centering 
    \begin{tabular}{c|c|c|c|c}
         \hline
         Priors & $a$ & $b$ & $C_0$ & $\sigma_{\rm int}$\\
         \hline
         F.P. Fit & $-0.830 \pm 0.158$   & $0.481 \pm 0.170$   & $25.807\pm8.978$   & $0.337\pm0.070$  \\
         \hline
         1$\sigma$ G. & $-0.830^{+0.130}_{-0.128}$ (15.7\%) & $0.305^{+0.094}_{-0.089}$ (30.8\%) & $34.808_{+4.591}^{-4.862}$ (14.0\%) & $0.441_{-0.055}^{+0.058}$ (13.2\%) \\
         2$\sigma$ G. & $ -0.860^{+0.231}_{-0.230}$ (26.9\%) & $0.210^{+0.097}_{-0.160}$ (76.2\%) & $ 40.981_{+8.109}^{-5.121}$ (19.8\%) & $0.561^{+0.101}_{-0.115}$ (20.5\%) \\
         3$\sigma$ G. & $ -0.863^{+0.304}_{-0.299}$ (35.2\%) & $<0.248$ & $ 41.548_{+8.178}^{-6.225}$ (19.7\%) & $0.667^{+0.136}_{-0.155}$ (20.4\%) \\
         5$\sigma$ G. & $-0.887^{+0.422}_{-0.403}$ (47.6\%) & $<0.250$ & $41.959_{+9.334}^{-6.287}$ (22.3\%) & $ 0.879^{+0.214}_{-0.252}$ (28.7\%) \\
         \hline 
         3$\sigma$ F. & $-0.890_{-0.299}^{+0.299}$ (33.7\%) & $0.030^{+0.225}_{-0.206}$ & $48.874_{+10.678}^{-11.610}$ (23.8\%) & $0.572^{+0.119}_{-0.053}$ (20.8\%) \\
         5$\sigma$ F. & $ -0.961_{-0.480}^{+0.451}$ (50.0\%) & $<0.218$ & $>41.575$ & $0.726^{+0.185}_{-0.095}$ (25.5\%) \\
         \hline
    \end{tabular}
}
\caption{Constraints on the parameters of interest for the 3D Dainotti relation assuming Gaussian (G.) and flat (F.) priors with an $n$-$\sigma$ progression with the platinum sample. We report the mean and associated 68\% C.L. for each parameter. The relative percentage changes in given in brackets.}
\label{tab:3d_calib}
\end{table}

\begin{table}[t]
{\renewcommand{\arraystretch}{1.5} \setlength{\tabcolsep}{25 pt} \centering 
    \begin{tabular}{c|c|c|c}
         \hline
         Priors & $a$ & $C_0$ & $\sigma_{\rm int}$\\
         \hline
         F.P. Fit & $-1.035 \pm 0.162$ & $51.157 \pm 0.535$ & $0.407\pm 0.081$\\
         \hline
         1$\sigma$ G. & $-1.036_{-0.112}^{+0.113}$~~ (10.9\%) &  $51.123_{-0.377}^{+0.371}$~~ (0.7\%) & $0.479_{-0.067}^{+0.069}$~~  (14.4\%) \\
         2$\sigma$ G. & $-1.030_{-0.206}^{+0.205}$~~ (20.0\%) & $ 51.094_{-0.675}^{+0.681}$~~ (1.3\%) & $0.592_{-0.114}^{+0.126}$~~ (21.3\%) \\
         3$\sigma$ G. & $-1.028_{-0.287}^{+0.288}$~~ (28.0\%) &  $51.086_{-0.947}^{+0.949}$~~ (1.9\%) & $0.709_{-0.159}^{+0.181}$~~ (25.5\%) \\
         5$\sigma$ G. & $-1.030_{-0.437}^{+0.423}$~~ (42.4\%) &  $51.100_{-1.401}^{+1.440}$~~ (2.8\%) & $0.930_{-0.247}^{+0.293}$~~ (31.5\%) \\
         \hline
         3$\sigma$ F. & $-1.024_{-0.278}^{+0.274}$~~ (27.2\%) & $51.080_{-0.911}^{+0.919}$~~ (1.8\%) & $ 0.654_{-0.130}^{+0.111}$~~ (19.9\%) \\
         5$\sigma$ F. & $ -1.021_{-0.413}^{+0.405}$~~ (40.5\%) &  $51.063_{-1.343}^{+1.365}$~~ (2.7\%) & $0.839_{-0.197}^{+0.171}$~~ (23.5\%) \\
         \hline
    \end{tabular}
}
\caption{Constraints on the parameters of interest for the 2D Dainotti relation assuming Gaussian (G.) and flat (F.) priors with an $n$-$\sigma$ progression. We report the mean and associated 68\% C.L. for each parameter. The relative percentage changes for their uncertainties given in brackets.}
    \label{tab:2d_calib}
\end{table}

\begin{figure}[t]
    \centering
    \includegraphics[width = 0.485\textwidth]{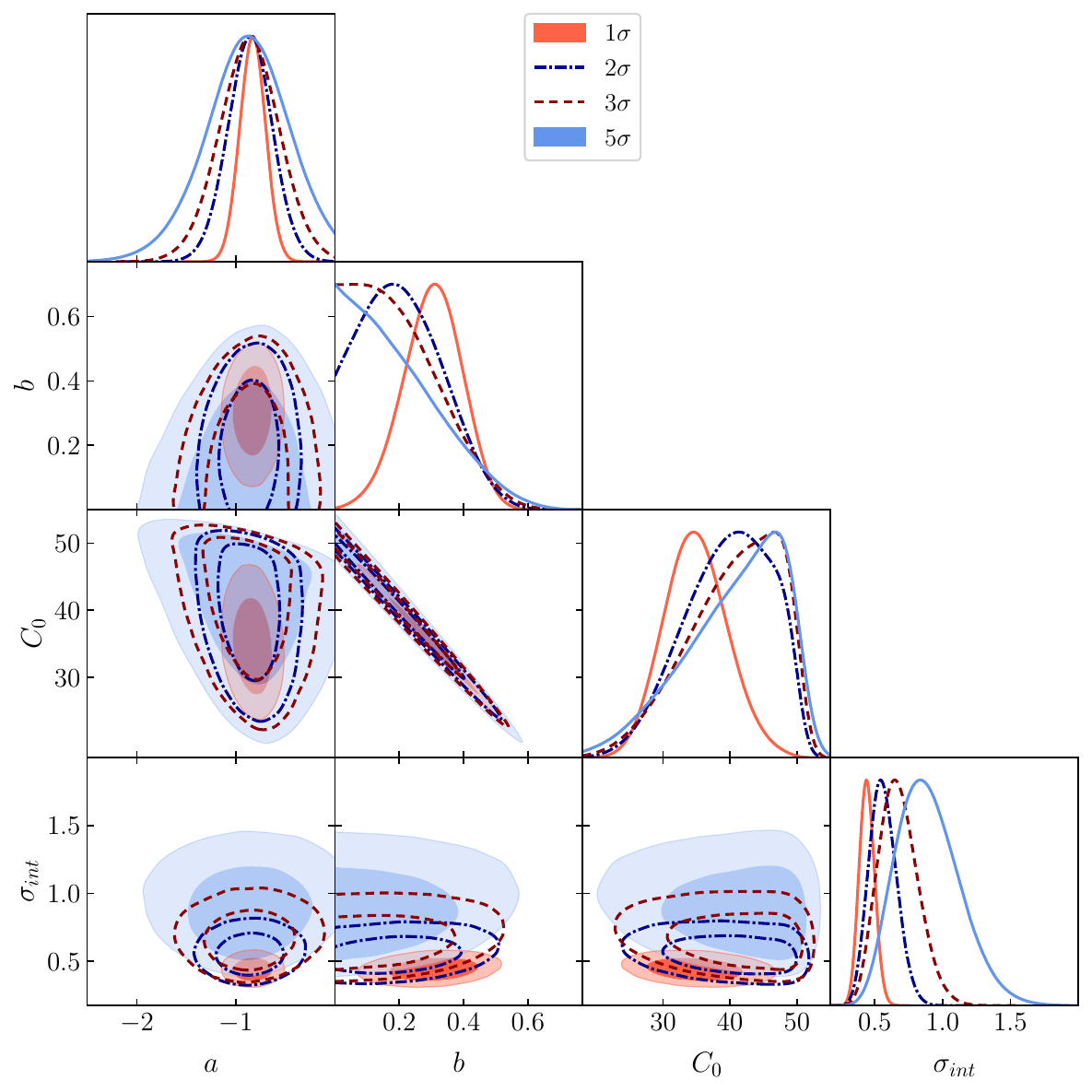}
    \includegraphics[width = 0.485\textwidth]{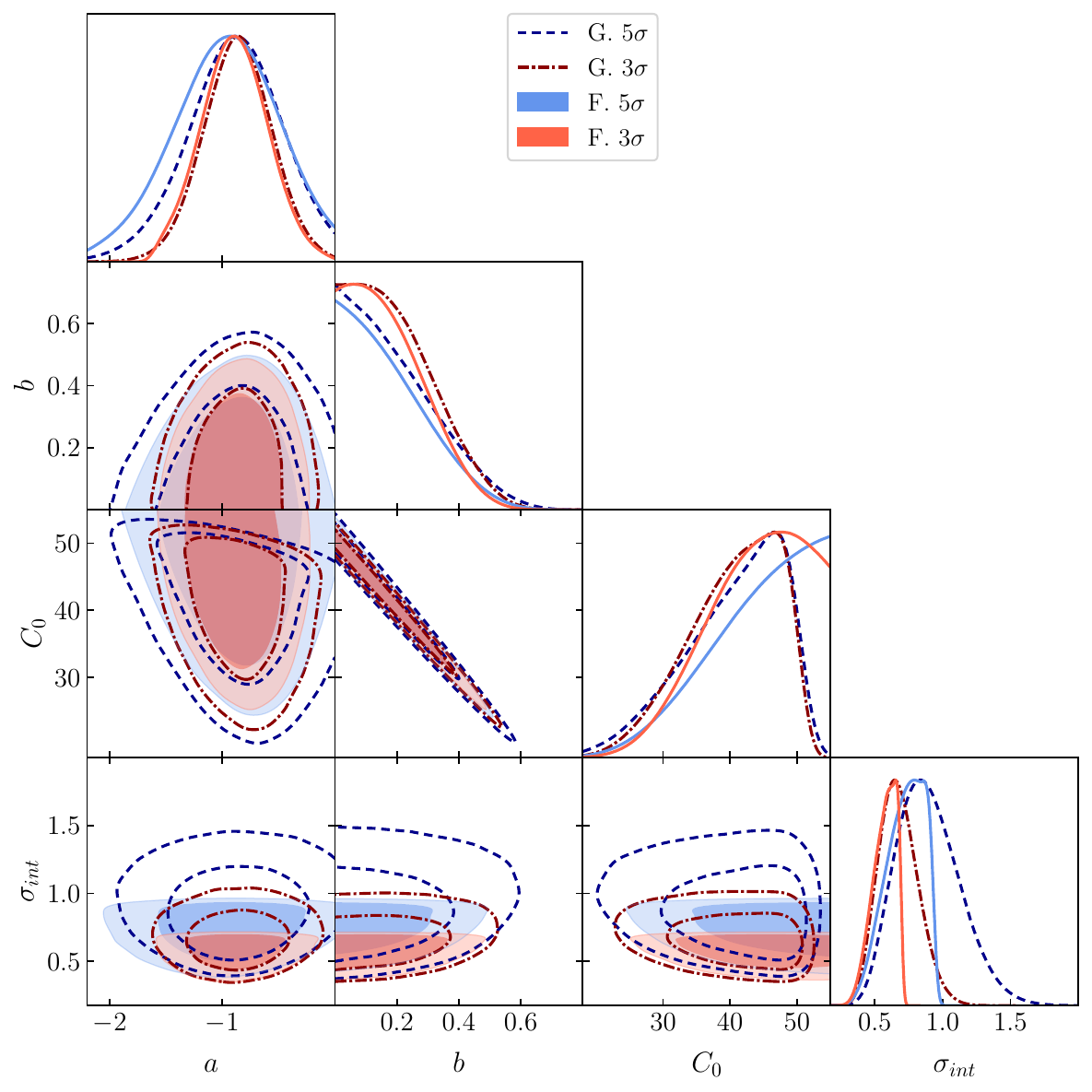}
\caption{1D posteriors and 2D contours at 68\% and 95\% confidence levels (C.L.) for the analysis of the 3D Dainotti relation. The analysis assumes Gaussian priors with an $n$-$\sigma$ progression and applies physical constraints $a < 0$, $b > 0$, and $\sigma_{\rm int} > 0$ in the left panel. Comparison using Gaussian (G.) priors and the corresponding Flat (F.) priors at 3$\sigma$ and 5$\sigma$ in the right panel.}
    \label{fig:3d_cal_mcmc}
\end{figure}

\begin{figure}[t]
    \centering
    \includegraphics[width = 0.45\textwidth]{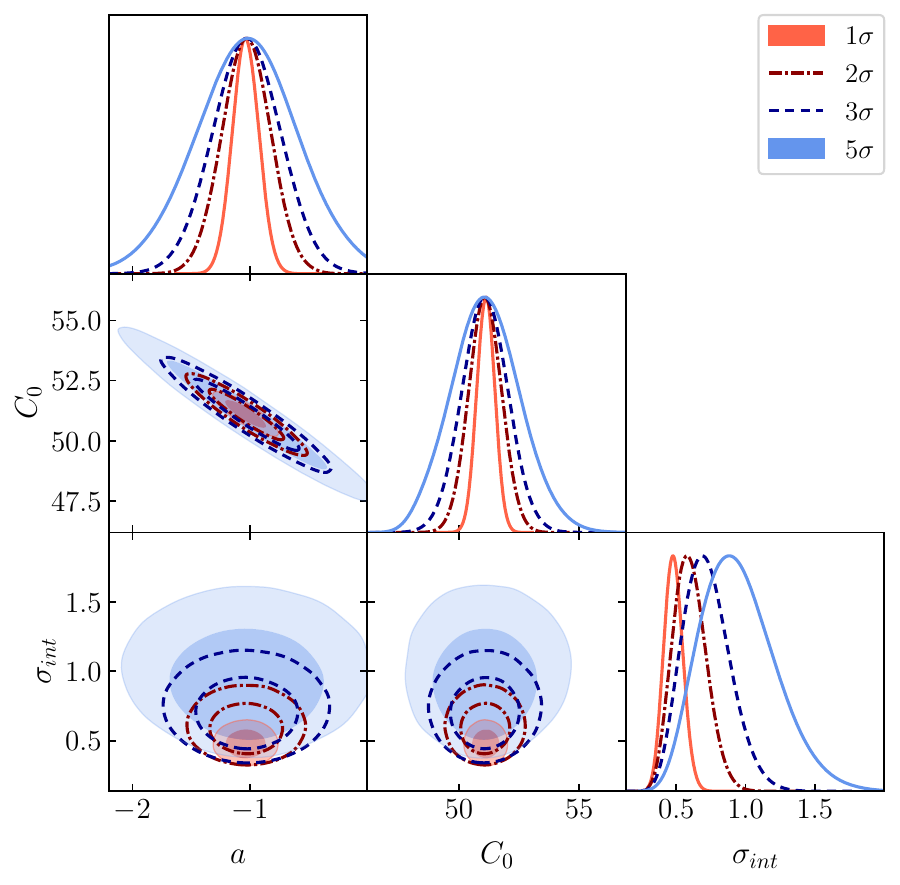}
    \includegraphics[width = 0.45\textwidth]{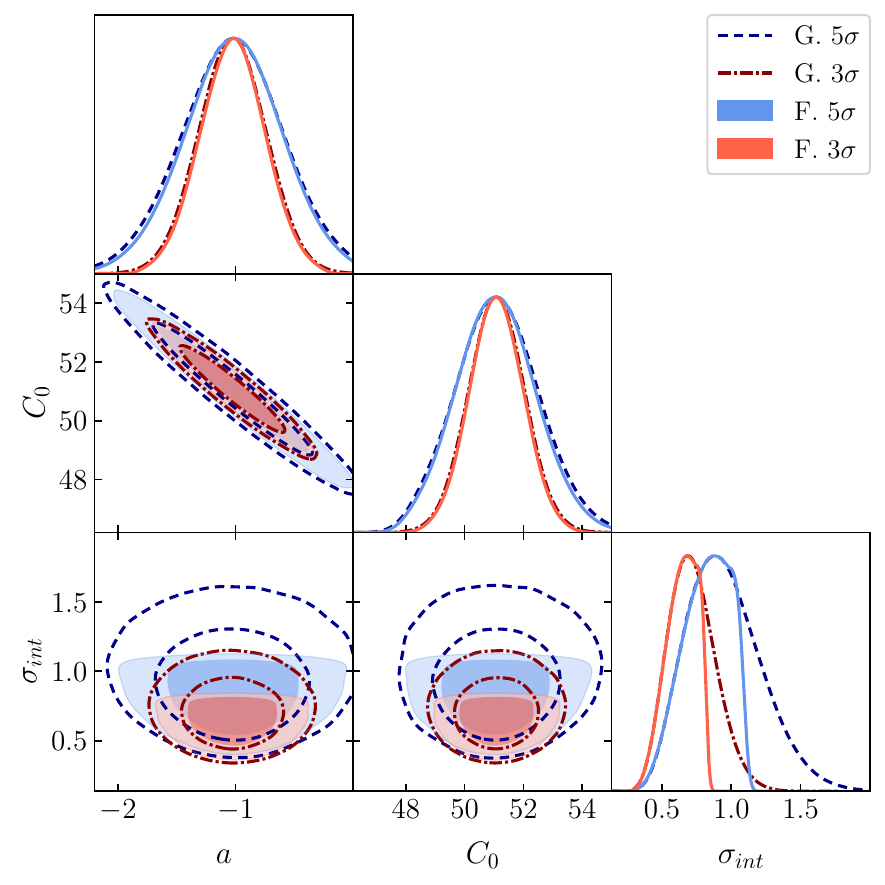}
\caption{1D posteriors and 2D contours at 68\% and 95\% confidence levels (C.L.) for the analysis of the 2D Dainotti relation. The analysis assumes Gaussian priors with an $n$-$\sigma$ progression and applies physical constraints $a < 0$, and $\sigma_{\rm int} > 0$ in the left panel. Comparison using Gaussian (G.) priors and the corresponding Flat (F.) priors at 3$\sigma$ and 5$\sigma$ in the right panel.}
    \label{fig:2d_cal_mcmc}
\end{figure}

\subsection{With fixed values for \texorpdfstring{$ \mathbf{\log T_X^{\star}}$}{1}, \texorpdfstring{$\mathbf{\log F_X}$}{2}, \texorpdfstring{$\mathbf{\log K_{\rm plateau}}$}{3}, \texorpdfstring{$\mathbf{\log F_{\rm peak}}$}{4}, \texorpdfstring{$\mathbf{\log K_{\rm prompt} }$}{5} }

The results of the MCMC analyses assuming fixed values of parameters $\log T_X^{*}$, $\log F_X$, $\log K_{\rm plateau}$, $\log F_{\rm peak}$, and $\log K_{\rm prompt}$ in FP-fit are shown in Fig.~\ref{fig:3d_cal_mcmc}, while the 3D Dainotti relation constraints on the calibration parameters are given in Table~\ref{tab:3d_calib}. Here, the Gaussian priors scenario are largely compatible with each other. This is particularly true for the case of the $3\sigma$ and $5\sigma$ scenarios, while the $1\sigma$ case did show statistical differences, particularly for the $b$ and $C_0$ parameters. This may be a result of the MCMC sampling being most limited for the $1\sigma$ scenario in terms of the sample of the parameter space. Another important feature of the Gaussian scenarios is that the $3\sigma$ and $5\sigma$ scenarios show a widening for the calibration parameters which gives less information. On the other hand, we sort of expect this outcome given also the results shown in Favale et al. \cite{Favale:2024lgp}.

The flat prior scenarios show relatively strong consistency with their Gaussian prior analogues, and are more coherent with each other. On the other hand, except for the $\sigma_{\rm int}$ parameter, these posteriors are wider than analogous Gaussian cases. 
On the other hand, the contours show more information on the covariances between the calibration parameters. For instance, the $b$ and $C_0$ parameters appear to be anti-correlated which occurs since $C_0$ emerges as a constant in the calibration equation, while $b$ relates the peak luminosity of the sources with the plateau luminosity. There is a general and pronounced reduction of the uncertainties on the $\sigma_{\rm int}$ parameter in relation to all the other parameters. The other parameters show reasonable independence.

In all cases, the numerical constraints are shown in Table~\ref{tab:3d_calib} where all five scenarios are shown for the calibration parameters $a$, $b$, $C_0$, and $\sigma_{\rm int}$. These are shown together with their uncertainties at $68\%$ C.L., which support the abovementioned descriptions. Here, the value of $a$ seems to prefer a value higher than $-1$ showing a shallower anti-correlation between $L_X$ and $T_X$ as compared with the literature, while $b$ shows a marginal preference for being positive, although we had already set the prior for $b>0$. 

This reinforces the need to consider the 2D Dainotti relation. $C_0$ normalizes the fundamental plane and gives a strong degeneracy with $b$. While we constrain $C_0$ to be larger than $0$, it also exhibits an upper bound due to the reduction in the uncertainties, which is connected to the lower bound of $b$ being limited to zero. For comparison of the calibration parameters, we next perform a similar analysis for the 2D Dainotti relation, where $b$ is 0.

The numerical constraints on the respective calibration parameters $a$, $C_0$, and $\sigma_{\rm int}$ along with their uncertainties at $68\%$ C.L. are outlined in Table~\ref{tab:2d_calib}. 
As can be clearly observed in Fig.~\ref{fig:3d_cal_mcmc}, there is a strong degeneracy between the parameters $C_0$ and $b$. Consequently, when fixing $b=0$, the derived uncertainties in the parameter $C_0$ are strongly suppressed. Indeed, Gaussian posterior distributions are inferred for the parameter constraints of $C_0$, as depicted in Fig.~\ref{fig:2d_cal_mcmc}. This degeneracy is then inherited by $a$ which is now strongly anti-correlated with $C_0$. Moreover, the reduction of the uncertainty in the $\sigma_{\rm int}$ parameter concerning all the other parameters is also observed in this case, however less pronounced with respect to the 3D Dainotti relation case.

\subsection{With \texorpdfstring{$\mathbf{\log T_X^{\star}}$}{1}, \texorpdfstring{$\mathbf{\log F_X}$}{2}, \texorpdfstring{$\mathbf{\log K_{\rm plateau}}$}{3}, \texorpdfstring{$\mathbf{\log F_{\rm peak}}$}{4}, \texorpdfstring{$\mathbf{\log K_{\rm prompt}}$}{5} as nuisance parameters}

\begin{table}[t]
{\renewcommand{\arraystretch}{1.5} \setlength{\tabcolsep}{10 pt} \centering 
    \begin{tabular}{c|c|c|c|c}
         \hline
         Nuisance Priors & $a$ & $b$ & $C_0$ & $\sigma_{\rm int}$\\
         \hline
         1$\sigma$ G. & $-0.786\pm 0.081$ (10.3\%) & $0.5005^{+0.0066}_{-0.0060}$ (1.2\%) & $24.640\pm 0.081$ (0.3\%) & $0.444^{+0.071}_{-0.084}$ (18.9\%) \\
         3$\sigma$ G. & $-0.72\pm 0.15$ (20.8\%) & $0.495\pm 0.010$  (2.0\%) & $24.71\pm 0.19$ (0.8\%) & $0.57^{+0.12}_{-0.15}$ (26.3\%) \\
         5$\sigma$ G. & $-0.65\pm 0.23$ (35.4\%) & $0.489^{+0.018}_{-0.017}$ (3.7\%) & $24.76\pm 0.31$ (1.3\%) & $ 0.70^{+0.16}_{-0.20}$ (28.6\%) \\
         \hline
    \end{tabular}
}
\caption{Constraints on the parameters governing the 3D Dainotti relation assuming flat priors for the GRB Calibration parameters with Gaussian (G.) priors on the nuisance parameters in $n$-$\sigma$ progression. We report the mean and associated 68\% C.L. for each parameter. The relative percentage changes in given in brackets.}
\label{tab:3d_calib_with_nuisance}
\end{table}

\begin{figure}[t]
    \centering
    \includegraphics[width = 0.325\textwidth]{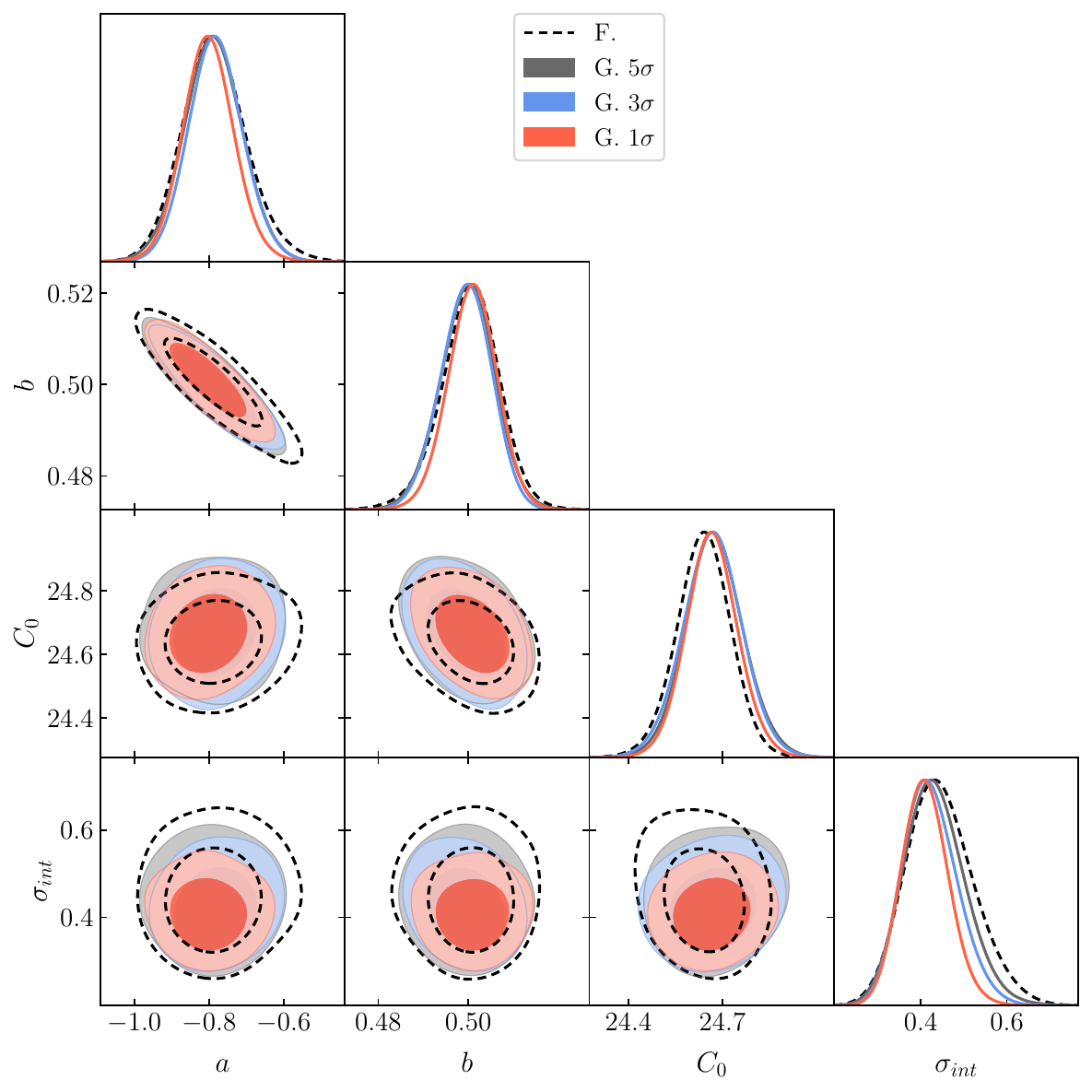}
    \includegraphics[width = 0.325\textwidth]{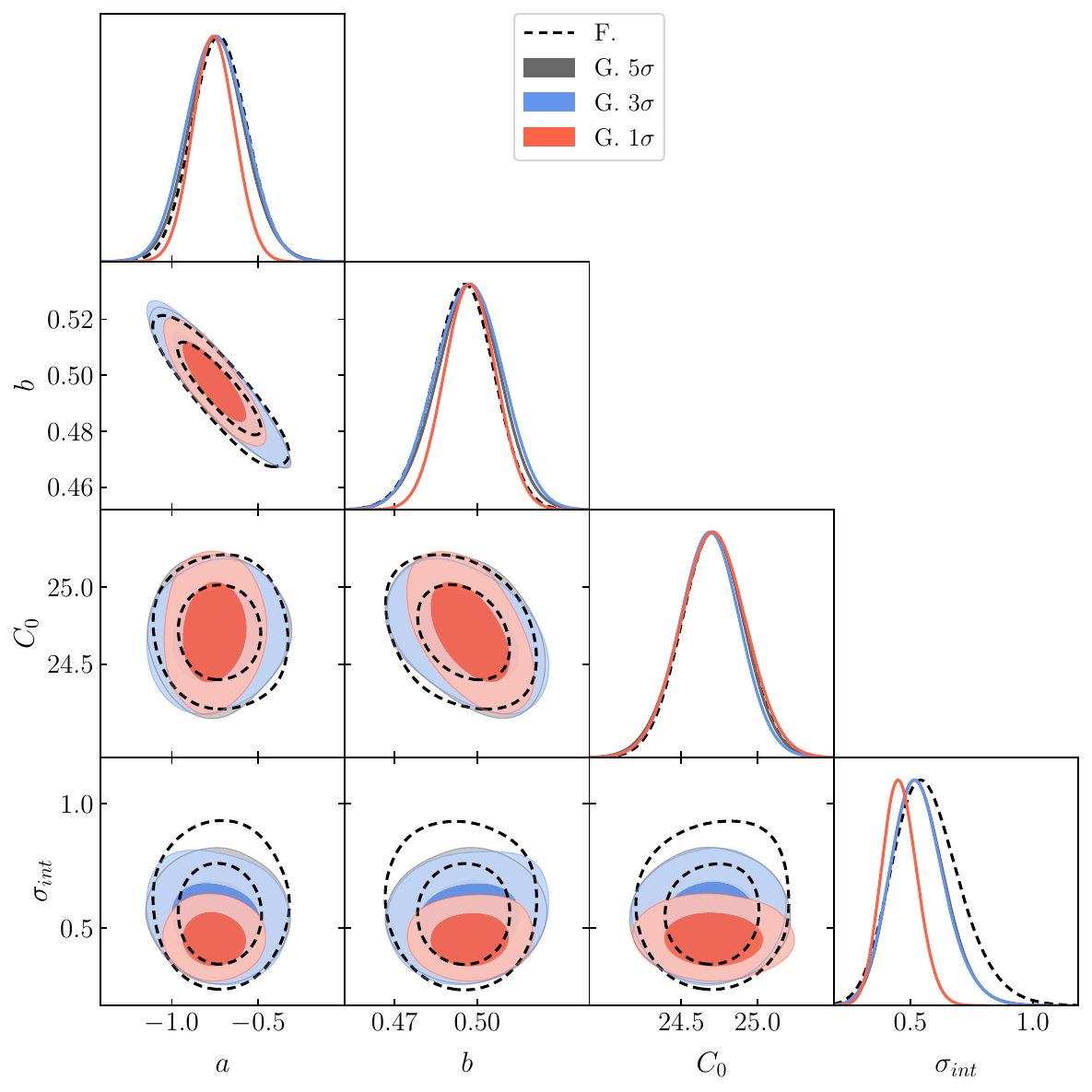}
    \includegraphics[width = 0.325\textwidth]{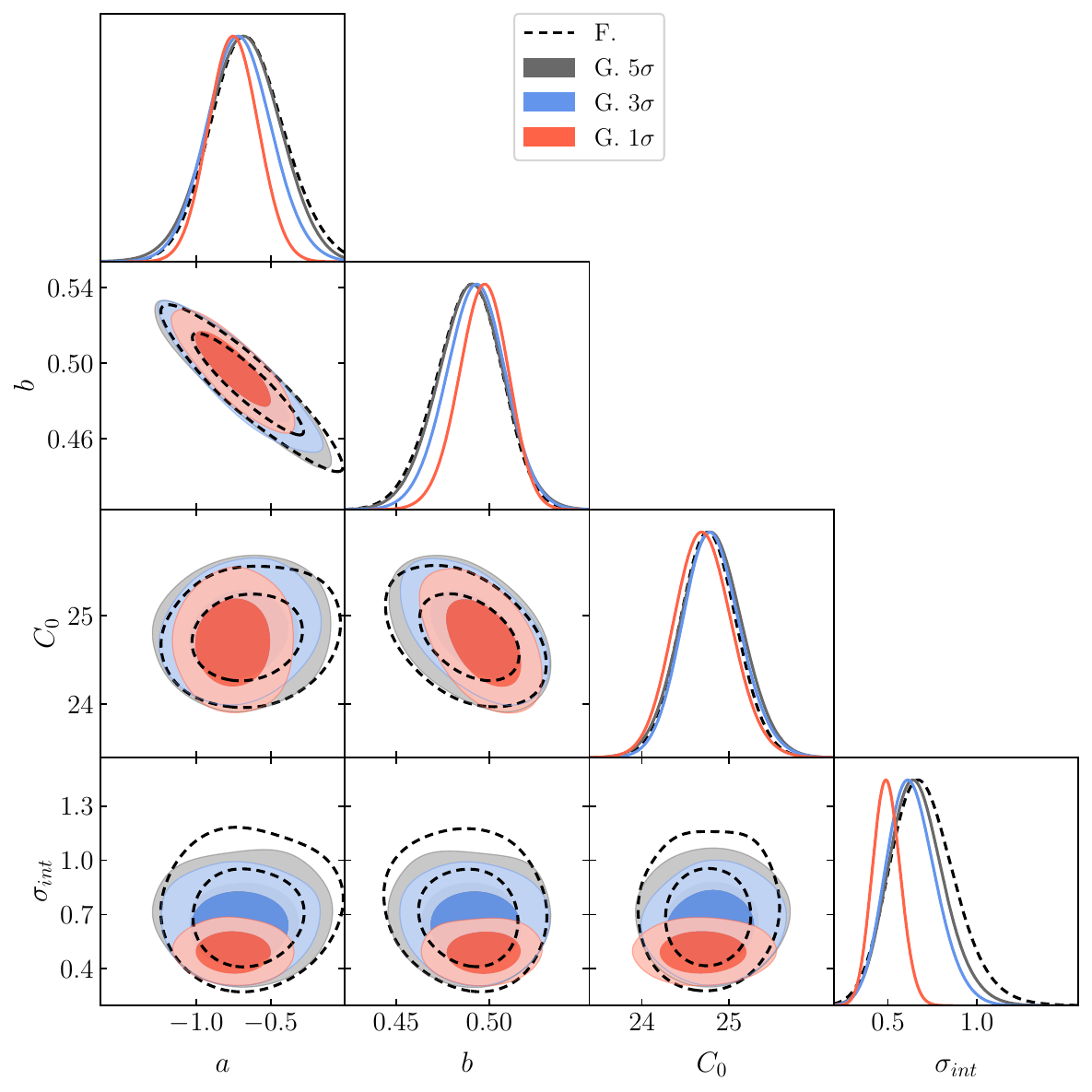}
\caption{1D posteriors and 2D contours at 68\% and 95\% confidence levels (C.L.) for the analysis of the 3D Dainotti relation on marginalizing over GRB nuisance parameters assuming $1\sigma$, 3$\sigma$, and 5$\sigma$ Gaussian priors.}
    \label{fig:3d_cal_mcmc_with_nuisance}
\end{figure}

We now proceed to constrain the parameters governing the 3D Dainotti relations, with the five parameters defining the F.P. fit (i.e., $\left\lbrace \log T_X^{\star}, \log F_X, \log K_{\rm plateau}, \log F_{\rm peak}, \log K_{\rm prompt} \right\rbrace$) treated as nuisance parameters in the MCMC analysis. We then marginalize over these nuisance parameters to obtain confidence intervals on the parameters of interest. Fig. \ref{fig:3d_cal_mcmc_with_nuisance} and Table \ref{tab:3d_calib_with_nuisance} illustrate the marginalized constraints on the calibration parameters while considering these nuisance variables. Notably, the results for the $3\sigma - 5\sigma$ Gaussian and Flat prior scenarios are mutually consistent, especially for the $b$ and $C_0$ parameters.

A striking feature emerges for the $b$ parameter, which is non-negative ($>0$). It appears to be tightly constrained at approximately $\sim 0.49$, a value consistent with the F.P. fit applied to the same GRB sample. This result provides precise confidence intervals on $C_0$, aligning closely with those obtained from the F.P. fit. However, we observe a slight increase (from 18\% to 28\%, depending on the priors) in the intrinsic scatter parameter $\sigma_{\rm int}$ when assuming flat priors for the GRB calibration parameters. This increase may be due to the broader range of values allowed for the nuisance parameters under flat priors, leading to greater uncertainty in the intrinsic scatter.  We expect these enlarged uncertainties from the construction of the nuisance parameters.

The differences in results between treating nuisance parameters as fixed versus allowing them to vary freely in the MCMC analysis can be largely attributed to how each approach constrains the parameter space. 
When nuisance parameters are fixed, the primary calibration parameters (e.g., $a$, $b$, $C_0$ and $\sigma_{\rm int}$) are limited to the current data sample.
Allowing nuisance parameters to vary introduces flexibility into the model, enabling it to account for a wider parameter space for the uncertainties and variability in the observations, also accounting for future observations that would resemble the current ones. This marginalization over nuisance parameters provides a more accurate representation of the data, leading to refined estimates of the GRB calibration parameters. Furthermore, the process of marginalization can help smooth out the effects of noise or outliers, resulting in more precise and reliable estimates of the key parameters of interest.

\section{Calibration of the 3D Dainotti relation including redshift evolution corrections} \label{sec:redshift_evo}

We have discussed that there is a shift in the parameters for the $a$ and $b$ compared to what it is expected in the literature \cite{Dainotti:2020azn}. However, one can notice that, indeed, in the more complex and reliable treatment, we need to consider the correction for selection biases due to the threshold truncation and to redshift evolution as shown in \cite{,Dainotti:2022jzv,10.1093/mnras/stac2752}. Notwithstanding the strength of these 2D and 3D correlations, one needs to correct for bias or selection effects: distortion of statistical correlations resulting from the method of collecting samples. Gathering data from a satellite with a given flux limit prevents us from seeing a representative sample of events. Thus, it is necessary to determine the true correlations among the variables, not those introduced by the observational biases. The common attempts to correct such effects \cite{1925MeLuF.106....1M} suffer from a major shortcoming: they are valid only if the variables are independent (or uncorrelated). GRB observables might be affected by redshift evolution \cite{Dainotti2013}; thus, a quantitative determination of this effect is needed \cite{Dainotti:2013fra,Dainotti:2013cta}. 

To overcome both the problems of the selection effects and redshift evolution \cite{Efron1992}, hereafter EP, suggested a technique that was successfully applied to GRBs. This method has been implemented from a two (luminosity-redshift space) to a three-dimensional space (luminosity, time, and redshift) by \cite{Dainotti:2013fra} and applied to an updated $L_X-T^{*}_X$ correlation for a sample of 101 GRBs. They determined the dependence of the luminosities and time distributions on the redshift, and they removed such dependence by creating new observables, de-evolved, divided by their evolution functions. This method demonstrated that the correlation is intrinsic at the 12$\sigma$ level, and thus, it can be used to constrain physical models of the plateau emission.  This gives more solid physical grounding to the application for this relation and its extension in 3D for cosmological purposes. Thus, if cosmological analyses are conducted without this correction, there will possibly be a shift of the parameters due to both selection biases and redshift evolution.

\begin{table}
{\renewcommand{\arraystretch}{1.5} \setlength{\tabcolsep}{3 pt} \centering   
    \begin{tabular}{c|c|c|c|c|c|c|c|c}
         \hline
         Priors & Sample & $a$ & $b$ & $C_0$ & $\sigma_{\rm int}$ & ${k_L}_x$ &  ${k_T^{\star}}_x$ & ${k_L}_{\rm peak}$ \\
         \hline
         F.P. Fit & P. & $-0.85 \pm 0.12$ & $0.49 \pm 0.13$ & $25.4 \pm 6.9$ & $0.18 \pm 0.09$ & $1.37^{+0.83}_{-0.93}$ & $-0.68^{+0.54}_{-0.82}$ & $0.44^{+1.37}_{-1.76}$ \\
          & W. &  $\cdots$ & $\cdots$ & $\cdots$ & $\cdots$ & $2.42^{+0.41}_{-0.74}$ & $-1.25^{+0.28}_{-0.27}$ &  $2.24 \pm 0.30$ \\
         \hline
         3$\sigma$ G. & P. & $-0.85\pm 0.27$ (31.8\%) & $0.225^{+0.094}_{-0.19}$ (84.4\%) & $39^{+9}_{-6}$ (23.0\%) & $0.71^{+0.14}_{-0.19}$ (26.7\%) & 1.37 & -0.68 & 0.44 \\
         3$\sigma$ G. & W. & $ -0.85\pm 0.26$ (30.6\%) & $0.217^{+0.085}_{-0.19}$ (87.5\%) & $ 39^{+8}_{-6}$ (21.0\%) & $0.70^{+0.14}_{-0.19}$ (27.1\%) & 2.42 & -1.25 & 2.24 \\
         5$\sigma$ G. & P. & $-0.89^{+0.42}_{-0.37}$ (47.2\%) & $< 0.284$ & $39^{+10}_{-6}$ (25.0\%) & $0.98^{+0.22}_{-0.32}$ (32.7\%) & 1.37 & -0.68 & 0.44 \\
         5$\sigma$ G. & W. & $ -0.87^{+0.41}_{-0.36}$ (47.1\%) & $< 0.273 $ & $39^{+9}_{-6}$ (23.0\%) & $0.96^{+0.22}_{-0.31}$ (32.3\%) & 2.42 & -1.25 & 2.24 \\
         \hline
         3$\sigma$ G. & P. & $-0.83\pm 0.26$ (31.3\%) & $0.227^{+0.082}_{-0.20}$ (88.1\%) & $38^{+8}_{-6}$ (21.0\%) & $0.73^{+0.16}_{-0.19}$  (26.0\%) & $1.4\pm 3.8$ & -0.68 & $2.1\pm 3.2$ \\
         3$\sigma$ G. & W. & $ -0.85\pm 0.27$ (31.8\%) & $0.212^{+0.069}_{-0.20}$ (94.3\%) & $ 39^{+8}_{-6}$ (21.0\%) & $0.69^{+0.15}_{-0.18}$  (26.1\%) & $2.4\pm 2.6$ & -1.25 & $2.29\pm 0.86$ \\
         5$\sigma$ G. & P. & $-0.86^{+0.45}_{-0.36}$ (52.3\%) & $< 0.279$ & $39^{+10}_{-7}$ (25.0\%) & $0.98^{+0.26}_{-0.30}$ (30.6\%) & $1.6\pm 6.1$ & -0.68 & $1.9\pm 4.7$ \\
         5$\sigma$ G. & W. & $ -0.86^{+0.44}_{-0.38}$ (51.2\%) & $< 0.276 $ & $39^{+9}_{-6}$ (23.0\%) & $0.97^{+0.25}_{-0.30}$ (30.9\%) & $2.2\pm 4.3$ & -1.25 & $2.2\pm 1.4$ \\
         \hline
         3$\sigma$ G. & P. & $-0.83\pm 0.27$ (32.5\%) & $0.227^{+0.085}_{-0.20}$ (88.1\%) & $38^{+8}_{-6}$ (21.0\%) & $0.73^{+0.16}_{-0.19}$  (26.0\%) & $1.4\pm 3.7$ & $-0.4\pm 2.7$ & $1.7\pm 3.9$ \\
         3$\sigma$ G. & W. & $ -0.85\pm 0.27$ (31.8\%) & $0.216^{+0.072}_{-0.20}$ (92.6\%) & $ 39^{+8}_{-6}$ (21.0\%) & $0.71^{+0.15}_{-0.18}$  (25.4\%) & $2.4\pm 2.5$ & $-1.2\pm 1.1$ & $2.32\pm 0.87$ \\
         5$\sigma$ G. & P. & $-0.86^{+0.44}_{-0.36}$ (51.2\%) & $< 0.267$ & $40^{+9}_{-6}$ (22.0\%) & $0.99^{+0.26}_{-0.31}$ (31.3\%) & $1.5\pm 6.2$ & $-0.4 \pm 4.4$ & $1.7\pm 6.3$ \\
         5$\sigma$ G. & W. & $ -0.86^{+0.45}_{-0.36}$ (52.3\%) & $< 0.262 $ & $40^{+9}_{-6}$ (22.0\%) & $0.98^{+0.25}_{-0.31}$ (31.6\%) & $2.4\pm 4.2$ & $-1.1 \pm 1.7$ & $2.3\pm 1.4$ \\
         \hline
    \end{tabular}
}
\caption{Constraints on the parameters of interest for the 3D Dainotti relation including redshift evolution corrections assuming Gaussian (G.) priors with an $n$-$\sigma$ progression. We report the mean and associated 68\% C.L. for each parameter. For the evolutionary coefficients, we use the results obtained with the full Platinum (P.) sample (50 GRBs) and the Whole (W.) sample (222 GRBs) in Dainotti et al. \cite{Dainotti:2023pwk}.}
\label{tab:evo_correction}
\end{table}

\begin{figure}
    \centering
    \includegraphics[width = 0.325\textwidth]{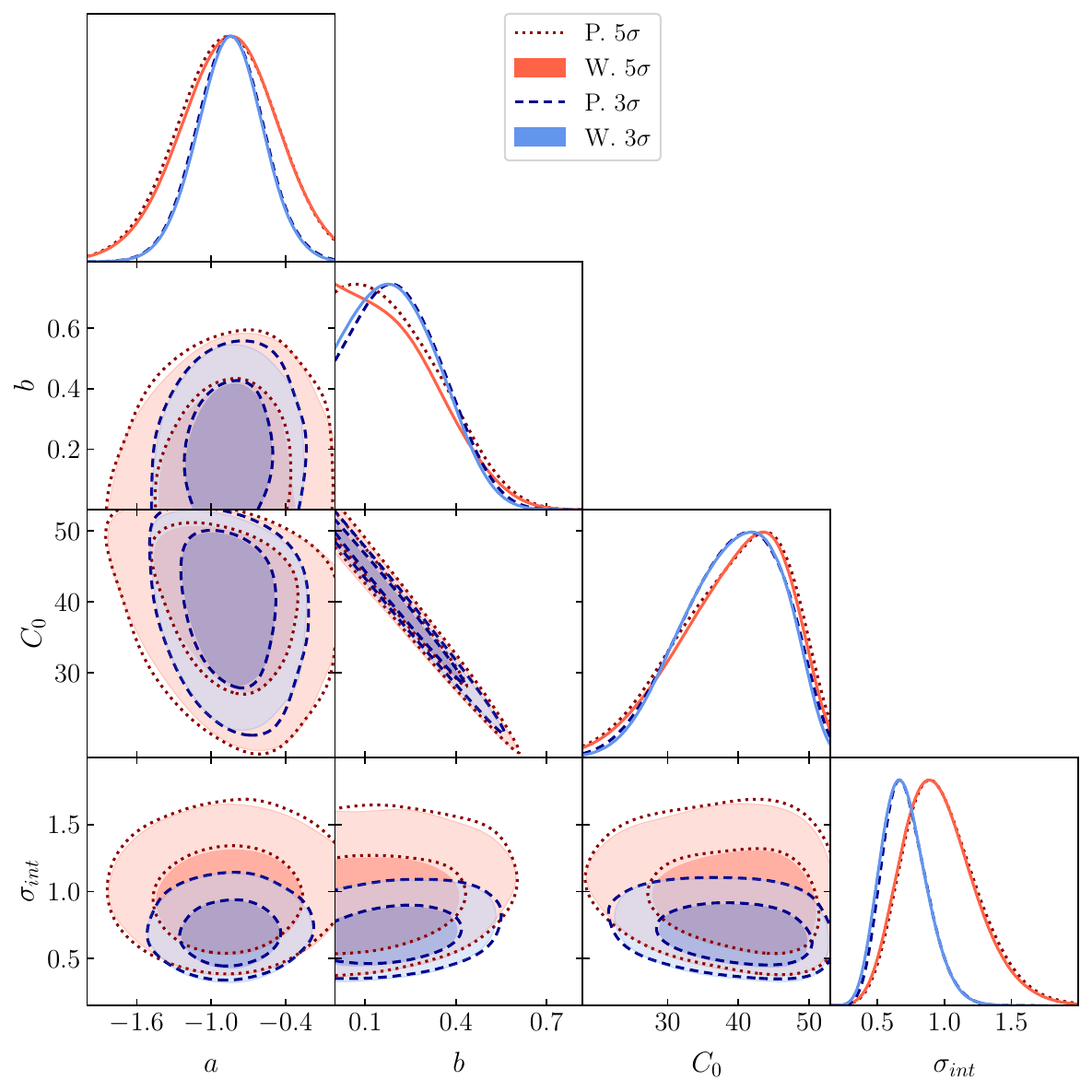}
    \includegraphics[width = 0.325\textwidth]{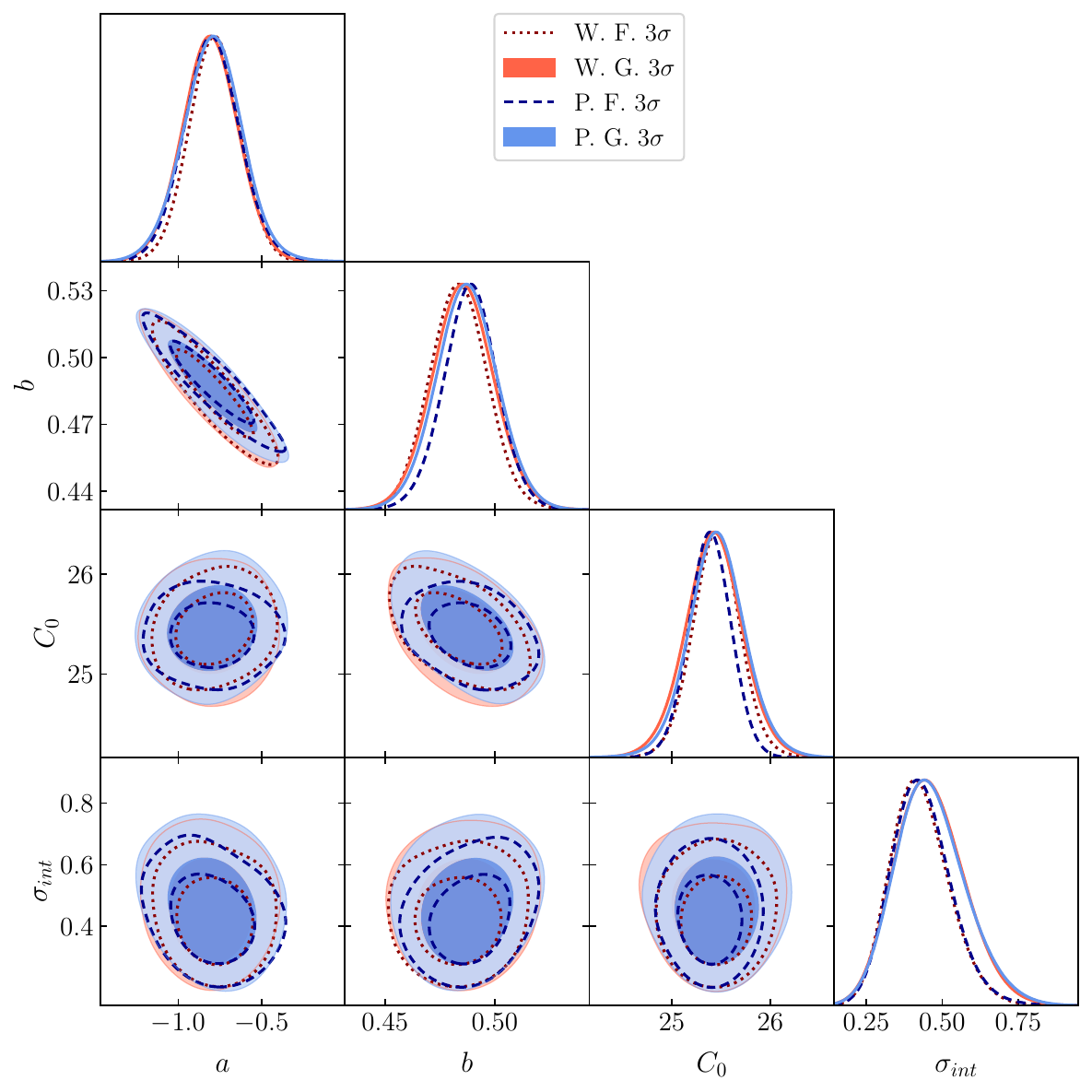}
    \includegraphics[width = 0.325\textwidth]{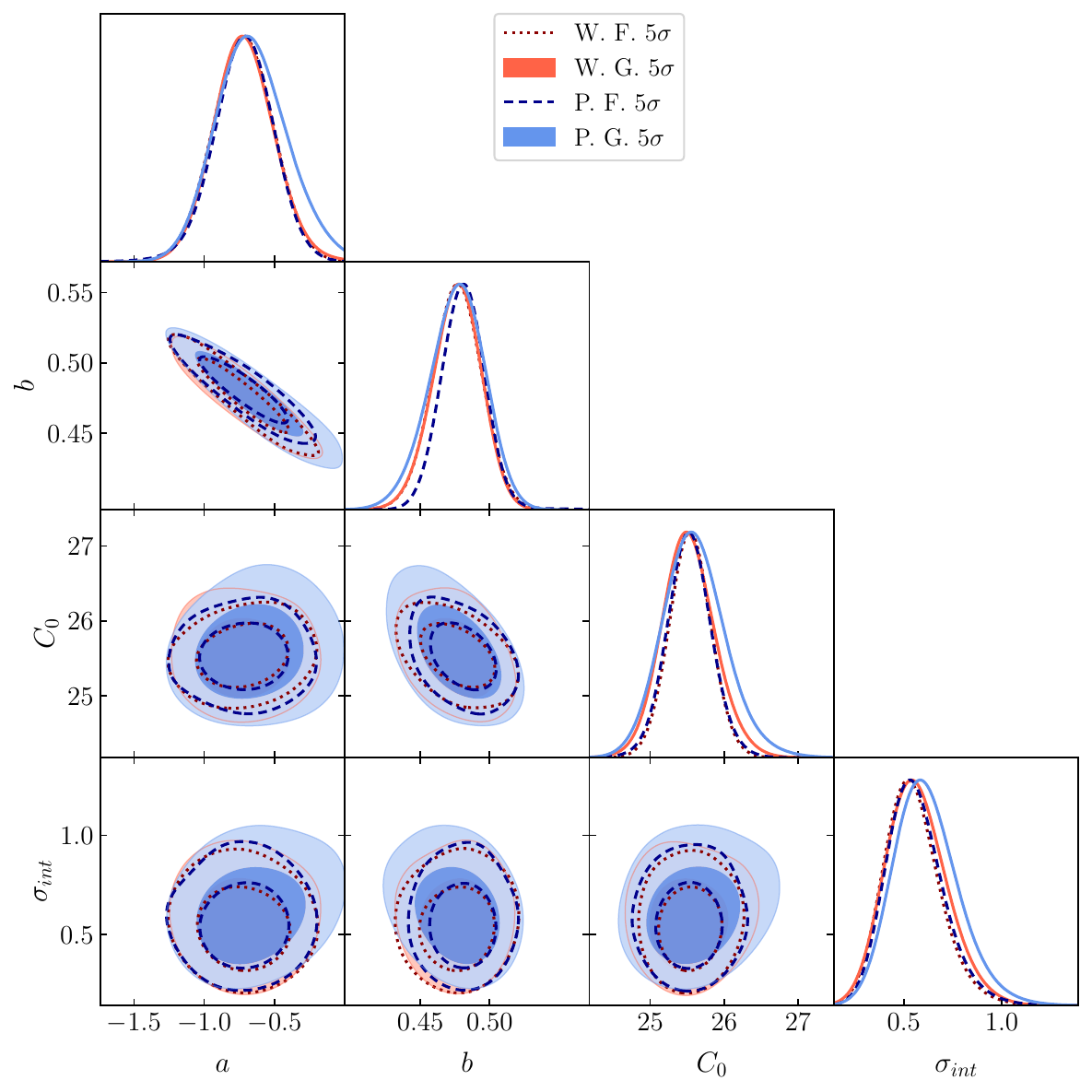}
    \caption{1D posteriors and 2D contour plots at 68\% and 95\% C.L. for GRB calibration parameters $a$, $b$, $C_0$ and $\sigma_{\rm int}$ including redshift evolution corrections in the 3D Dainotti relation, assuming the fixed values of the evolutionary coefficients ${k_L}_x$, ${k_T^{\star}}_x$ and ${k_L}_{\rm peak}$, sampled assuming Gaussian (in G.) and Flat (in F.) priors at 3$\sigma$ and $5\sigma$ from the Platinum sample (in P.) or Whole sample (in W.), on fixing the GRB nuisance parameters in left panel, with 3$\sigma$ and 5$\sigma$ marginalization over the nuisance parameters in middle and right panels. The relative percentage changes for their uncertainties are given in brackets.}
    \label{fig:k_fixed}
\end{figure}

\begin{table}
{\renewcommand{\arraystretch}{1.5} \setlength{\tabcolsep}{4 pt} \centering   
    \begin{tabular}{c|c|c|c|c|c|c}
         \hline
         Sample & Nuisance & Calibration  & $a$ & $b$ & $C_0$ & $\sigma_{\rm int}$ \\
         \hline
         P. & 1$\sigma$ G. & 3$\sigma$ G. & $-0.829\pm 0.077$ (9.3\%) & $0.4920\pm 0.0061$ (1.2\%) & $25.38\pm 0.11$ (0.4\%) & $0.316^{+0.057}_{-0.068}$  (28.1\%) \\
            &  & 3$\sigma$ F. & $-0.804\pm 0.073$ (9.1\%) & $0.4897\pm 0.0054$ (1.1\%) & $25.395\pm 0.086$ (0.3\%) & $0.290^{+0.043}_{-0.050}$  (22.7\%) \\
            &   &  5$\sigma$ G. & $-0.816\pm 0.084$ (10.3\%) & $0.4902\pm 0.0065$ (1.3\%) & $25.40\pm 0.11$ (0.4\%) & $0.310^{+0.055}_{-0.066}$  (27.7\%) \\
            &   & 5$\sigma$ F. & $-0.804\pm 0.073$ (9.1\%) & $0.4897\pm 0.0054$ (1.1\%) & $25.395\pm 0.086$ (0.3\%) & $0.290^{+0.043}_{-0.050}$  (22.7\%) \\
            \hline 
         W. & 1$\sigma$ G. & 3$\sigma$ G. & $-0.866\pm 0.077$ (8.9\%) & $0.4903\pm 0.0062$ (1.3\%) & $25.40\pm 0.11$ (0.4\%) & $0.305^{+0.052}_{-0.061}$  (20.0\%) \\
            &  & 3$\sigma$ F. & $-0.853\pm 0.061$ (7.2\%) & $0.4895\pm 0.0054$ (1.1\%) & $25.404^{+0.093}_{-0.084}$ (0.4\%) & $0.294^{+0.046}_{-0.055}$  (18.7\%) \\
            &   &  5$\sigma$ G. & $-0.865^{+0.080}_{-0.072}$ (12.4\%) & $0.4906^{+0.0066}_{-0.0073}$ (2.0\%) & $25.39^{+0.12}_{-0.11}$ (0.5\%) & $0.314^{+0.058}_{-0.072}$ (22.9\%) \\
            &   & 5$\sigma$ F. & $-0.853\pm 0.061$ (7.2\%) & $0.4895\pm 0.0054$ (1.1\%) & $25.404^{+0.093}_{-0.084}$ (0.4\%) & $0.294^{+0.046}_{-0.055}$  (18.7\%) \\
         \hline 
                  P. & 3$\sigma$ G. & 3$\sigma$ G. & $-0.79\pm 0.17$ (21.5\%) & $0.487\pm 0.013$ (2.7\%) & $25.46\pm 0.27$ (1.1\%) & $0.46^{+0.10}_{-0.12}$ (26.1\%) \\
            &  & 3$\sigma$ F. & $-0.80\pm 0.16$ (20.0\%) & $0.489\pm 0.012$ (2.5\%) & $25.39\pm 0.20$ (0.8\%) & $0.430^{+0.087}_{-0.10}$ (23.3\%) \\
            \hline 
                  W. & 3$\sigma$ G. & 3$\sigma$ G. & $-0.81\pm 0.16$ (19.8\%) & $0.486\pm 0.013$ (2.7\%) & $25.43\pm 0.27$ (1.1\%) & $0.46^{+0.10}_{-0.12}$ (26.1\%) \\
            &  & 3$\sigma$ F. & $-0.78\pm 0.14 $ (18.0\%) & $0.484\pm 0.012$ (2.5\%) & $25.45\pm 0.23$ (0.9\%) & $0.426^{+0.086}_{-0.10}$ (23.5\%) \\
         \hline 
                  P. & 5$\sigma$ G. & 5$\sigma$ G. & $-0.67^{+0.24}_{-0.27}$ (35.8\%) & $0.477^{+0.021}_{-0.019}$ (4.2\%) & $25.59^{+0.39}_{-0.44}$ (1.7\%) & $0.61^{+0.15}_{-0.18}$  (29.5\%) \\
            &  & 5$\sigma$ F. & $-0.72\pm 0.20$ (27.8\%) & $0.481\pm 0.015$ (3.1\%) & $25.53\pm 0.28$ (1.1\%) & $0.56^{+0.13}_{-0.16}$ (28.6\%) \\
            \hline 
                  W. & 5$\sigma$ G. & 3$\sigma$ G. & $-0.72\pm 0.20 $ (27.8\%) & $0.477\pm 0.016$ (3.4\%) & $25.52\pm 0.34$ (1.3\%) & $0.56^{+0.14}_{-0.17}$ (30.4\%) \\
            &  & 5$\sigma$ F. & $-0.72\pm 0.20$ (27.8\%) & $0.477\pm 0.017$ (3.6\%) & $25.54\pm 0.27$ (1.1\%) & $0.54^{+0.13}_{-0.15}$ (27.8\%) \\
         \hline
    \end{tabular}
}
\caption{Constraints on the parameters of interest for the 3D Dainotti relation including redshift evolution corrections, assuming Gaussian (G.) \& flat (F.) priors for the GRB Calibration parameters, along with Gaussian (G.) priors on the nuisance parameters in $n$-$\sigma$ progression. We report the mean and associated 68\% C.L. for each parameter. The evolutionary coefficients, are kept fixed to the results obtained with the full Platinum (P.) sample (50 GRBs) and the Whole (W.) sample (222 GRBs) in Dainotti et al. \cite{Dainotti:2023pwk}. The relative percentage of errors on each parameter is given in first brackets for comparison.}
\label{tab:evo_correction_marginalized}
\end{table}

On introducing this correction, the expression for the GRB fundamental plane takes the following form: \begin{equation} \label{eq:evo}
    \log L_X - {k_L}_x \log (1+z) = a (\log T^\star_X - k_{T^\star_x} \log(1+z)) + b (\log L_{\rm peak} - k_{L_{\rm peak}} \log(1+z)) + C_0 \, , 
\end{equation}
where ${k_L}_x$,  ${k_T^{\star}}_x$ and ${k_L}_{\rm peak}$ represent the evolutionary coefficients related to each physical feature. Similar to Eq. \eqref{eq: fc} \& \eqref{eq:dl_theo} we can now arrive at the relevant equation for the theoretical logarithmic-luminosity distance starting form Eq. \eqref{eq:evo}, and derive the new likelihood, which can account for redshift evolution effects and assess how it affects the correlation parameters. 

Here, we follow two main procedures: either use fixed evolution, in which the parameters of cosmology are used, or contemporaneously leave these parameters free to vary with the calibration parameters. We use both approaches to check the consistency between the two results, adopting Gaussian priors with a standard deviation equal to 1$\sigma$, 3$\sigma$, and 5$\sigma$ of the evolution values obtained with a standard flat $\Lambda$CDM model from the analysis of the Platinum sample reported in Dainotti et al. \cite{Dainotti:2023pwk}, i.e., $a = -0.85 \pm 0.12$, $b = 0.49 \pm 0.13$, $C_{0} = 25.4 \pm 6.9$ and $\sigma = 0.18 \pm 0.09$. For the evolutionary coefficients used in Eq. \eqref{eq:evo}, we leverage two sets of results: (a) Platinum sample of 50 GRBs, and (b) Whole sample of 222 GRBs, both detailed in \cite{Dainotti:2023pwk}. The corresponding values at 68\% C.L. are as follows: (a) ${k_L}_x$ =  $1.37^{+0.83}_{-0.93}$, ${k_T^{\star}}_x$ = $-0.68^{+0.54}_{-0.82}$ and ${k_L}_{\rm peak}$ =  $0.44^{+1.37}_{-1.76}$; (b) ${k_L}_x$ = $2.42^{+0.41}_{-0.74}$, ${k_T^{\star}}_x$ = $-1.25^{+0.28}_{-0.27}$ and ${k_L}_{\rm peak}$ =  $2.24 \pm 0.30$, respectively.

Table \ref{tab:evo_correction} summarizes the constraints on the GRB parameters of interest governing the 3D Dainotti relation including redshift evolution corrections for the fixed values of the evolutionary coefficients ${k_L}_x$, ${k_T^{\star}}_x$ and ${k_L}_{\rm peak}$, sampled assuming Gaussian (in G.) and Flat (in F.) priors at 3$\sigma$ and $5\sigma$ from the Platinum sample (in P.) or Whole sample (in W.) with an $n$-$\sigma$ progression. We report the mean and associated 68\% C.L. for each parameter. The left panel of Fig. \ref{fig:k_fixed} shows the obtained contour plots for this case. Our results indicate the preference for non-zero values of $b~ (>0)$, when compared to the case (see Fig. \ref{fig:3d_cal_mcmc} and Table \ref{tab:3d_calib}) when no redshift corrections were incorporated to the 3D Dainotti relation. During this exercise, the nuisance parameters defining the F.P. fit are initially kept fixed to their best-fit values. 

In the following step, we invoke this assumption and freely vary the nuisance parameters (i.e. $\log T_X^{\star}$, $\log F_X$, $\log K_{\rm plateau}$, $\log F_{\rm peak}$, $\log K_{\rm prompt}$) together with the GRB calibration parameters with redshift evolution correction terms. The results obtained from 3$\sigma$ and 5$\sigma$ marginalization over the nuisance parameters, given in Table \ref{tab:evo_correction_marginalized}, are illustrated in the middle and right panels of Fig. \ref{fig:k_fixed}. We find that the value of $\sigma_{\rm int}$ has slightly decreased, in comparison to Fig. \ref{fig:3d_cal_mcmc_with_nuisance}, when the redshift correction terms are included in the 3D Dainotti relation while GRB calibration. Moreover, the latter plots demonstrate that the Flat prior cases provide slightly better constraints on the GRB calibration parameters than those with Gaussian priors ones. The corresponding percentage changes for the parameters of interest with respect to their central values are also provided for comparison in brackets. 
Between the different tables (\ref{tab:3d_calib} and \ref{tab:evo_correction_marginalized})  there is a percentage decrease of 5\% for the slope, $a$, value from the 50\% Flat priors to 47.6\% for Gaussian priors when we do not consider redshift evolution at 5 $\sigma$ level, which is the confidence interval which carries the major difference. If we however, consider the same slope, $a$, but corrected for selection biases and redshift evolution we have a percentage change of  28.7\% (from $35.8 \%$ in the Gaussian prior to $27.8 \%$ for the flat priors) in the flat priors compared to the Gaussian. There is no particular trend of decrease between Gaussian and Flat priors, as is expected. It is worth noting that with the correction for evolution, the uncertainties decrease by 80\% for the flat priors and 25\% for Gaussian priors.

\begin{table}
{\renewcommand{\arraystretch}{1.35} \setlength{\tabcolsep}{9 pt} \centering   
    \begin{tabular}{c|c|c|c|c|c}
         \hline
         Sample & Nuisance & Calibration  & ${k_L}_x$ &  ${k_T^{\star}}_x$ & ${k_L}_{\rm peak}$ \\
         \hline
            P. & 3$\sigma$ G. & 3$\sigma$ F. & $1.37\pm 0.28$ & $-0.68$ & $0.53\pm 0.28$  \\
            &  & 3$\sigma$ F. & $1.38\pm 0.26$ & $-0.70\pm 0.22$ & $0.46\pm 0.23$  \\
            \hline 
           W. & 3$\sigma$ G. & 3$\sigma$ F. & $2.35\pm 0.26$ & $-1.25$ & $2.40\pm 0.23$  \\
            &  & 3$\sigma$ F. & $2.52^{+0.26}_{-0.29}$ & $-1.28\pm 0.23$ & $2.20\pm 0.23$  \\
            \hline 
           P. & 5$\sigma$ G. & 5$\sigma$ F. & $1.17^{+0.51}_{-0.43}$ & $-0.68$ & $0.60\pm 0.40$  \\
            &  & 5$\sigma$ F. & $1.42\pm 0.38$ & $-0.73\pm 0.40$ & $0.51\pm 0.38$  \\
            \hline 
           W. & 5$\sigma$ G. & 5$\sigma$ F. & $2.32\pm 0.36$ & $-1.25$ & $2.47\pm 0.42$  \\
            &  & 5$\sigma$ F. & $2.48\pm 0.40$ & $-1.34\pm 0.35$ & $2.11^{+0.37}_{-0.33}$  \\
            \hline 
    \end{tabular}
}
\caption{Constraints on redshift evolutionary coefficients, ${k_L}_x$, ${k_T^{\star}}_x$ and ${k_L}_{\rm peak}$, sampled from the Platinum (P.) sample (50 GRBs) and Whole (W.) sample (222 GRBs) assuming flat (F.) priors in $n$-$\sigma$ progression \cite{Dainotti:2023pwk}. We report the mean and associated 68\% C.L. for each parameter. }
\label{tab:evo_correction_result}
\end{table}

\begin{figure}
    \centering
    \includegraphics[width = 0.285\textwidth]{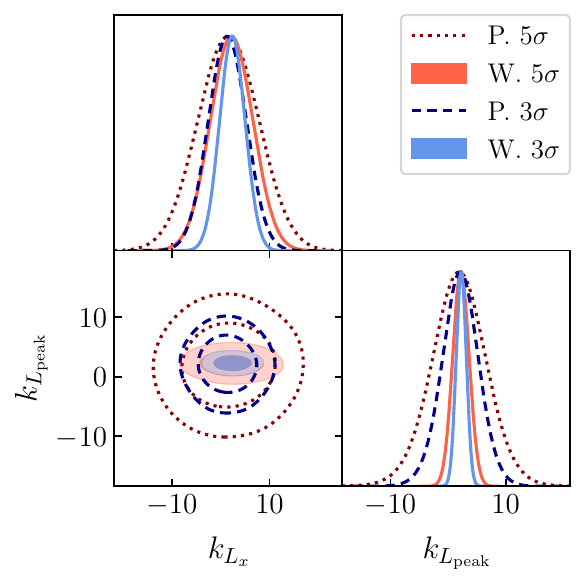} \hfill
    \includegraphics[width = 0.285\textwidth]{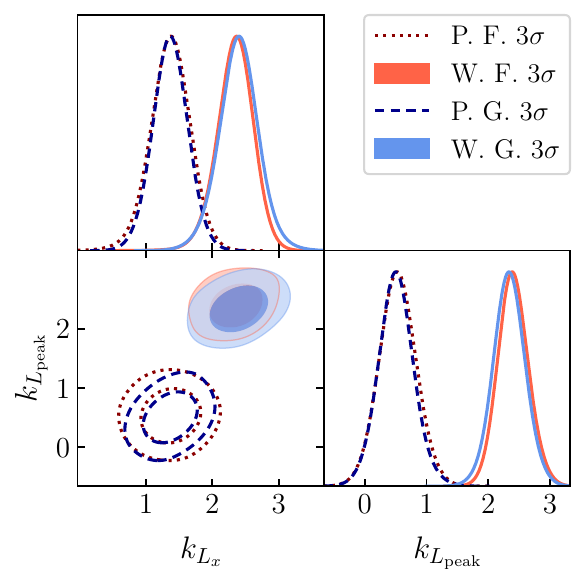} \hfill 
    \includegraphics[width = 0.285\textwidth]{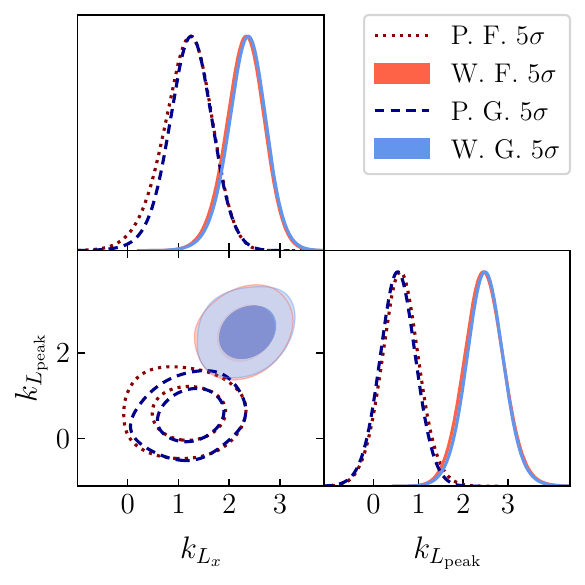}
    \caption{1D posteriors and 2D contour plots at 68\% and 95\% C.L. for GRB the evolutionary coefficients ${k_L}_x$ and ${k_L}_{\rm peak}$ keeping ${k_T^{\star}}_x$ fixed, considering redshift evolution corrections in the 3D Dainotti relation sampled assuming Gaussian (in G.) and Flat (in F.) priors at 3$\sigma$ and $5\sigma$ from the Platinum sample (in P.) or Whole sample (in W.), on fixing the GRB nuisance parameters in left panel, with 3$\sigma$ and 5$\sigma$ marginalization over the nuisance parameters in middle and right panels.}
    \label{fig:kt_fixed}
\end{figure}

\begin{figure}
    \centering
    \includegraphics[width = 0.325\textwidth]{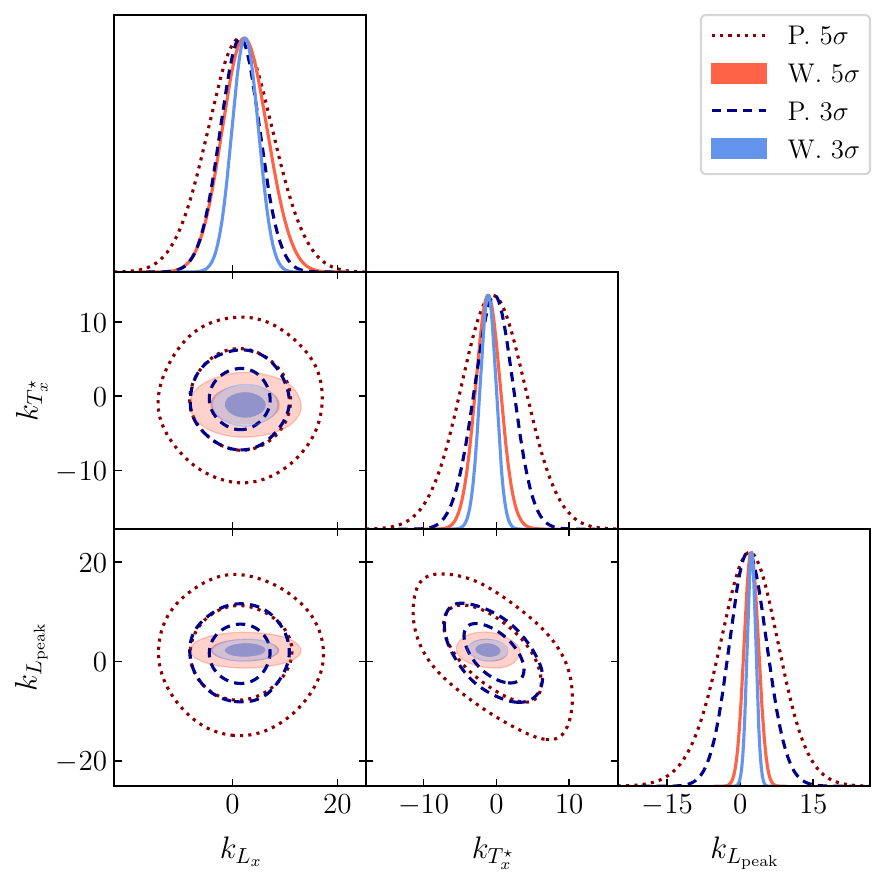}
    \includegraphics[width = 0.325\textwidth]{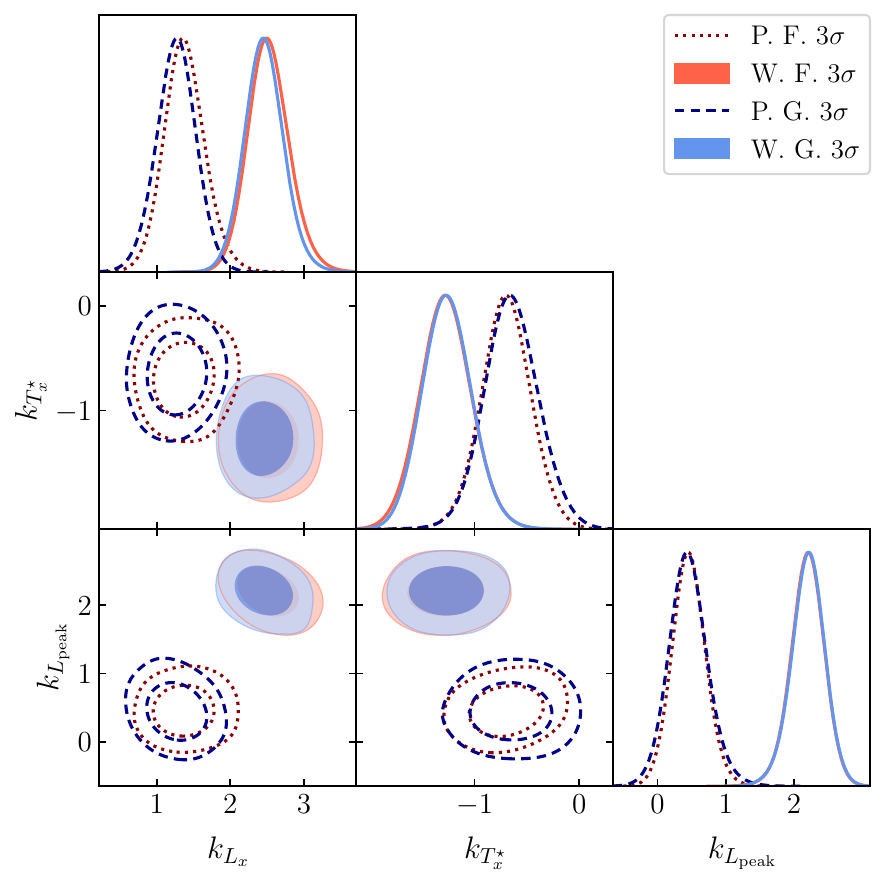}
 \includegraphics[width = 0.325\textwidth]{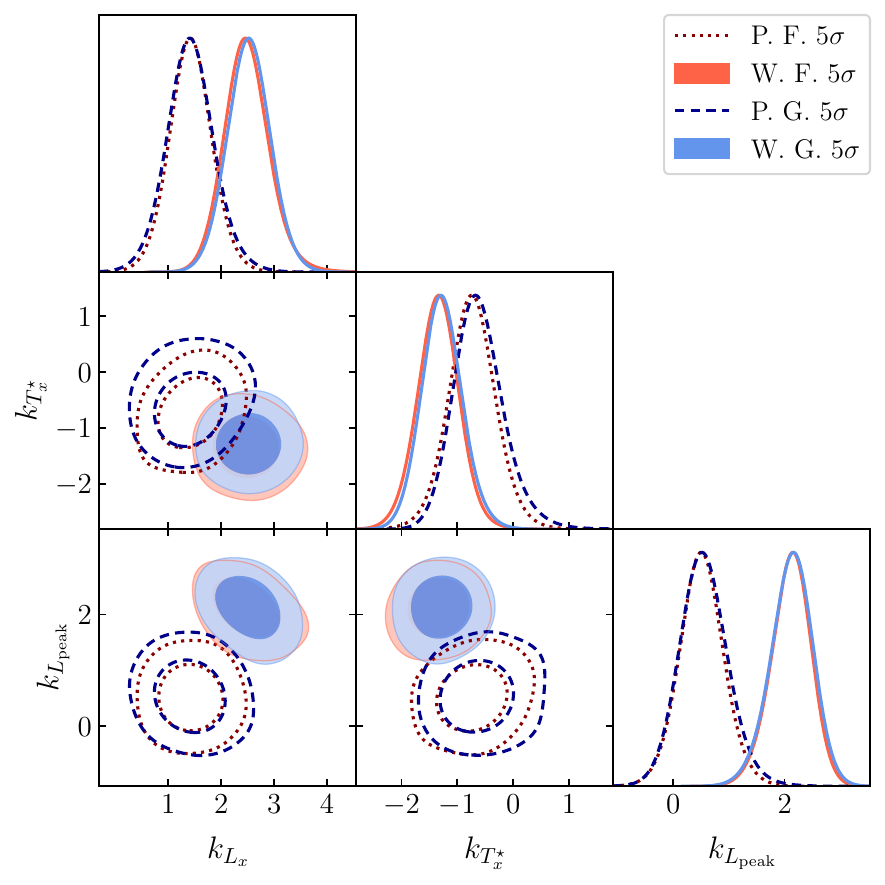}
    \caption{1D posteriors and 2D contour plots at 68\% and 95\% C.L. for GRB the evolutionary coefficients ${k_L}_x$, ${k_T^{\star}}_x$ and ${k_L}_{\rm peak}$, considering redshift evolution corrections in the 3D Dainotti relation sampled assuming Gaussian (in G.) and Flat (in F.) priors at 3$\sigma$ and $5\sigma$ from the Platinum sample (in P.) or Whole sample (in W.), on fixing the GRB nuisance parameters in left panel, with 3$\sigma$ and 5$\sigma$ marginalization over the nuisance parameters in middle and right panels.}
    \label{fig:all_k_free}
\end{figure}

The scatter reduction for $\sigma_{\rm int}$ goes from 0.726 in the case of no-correction for evolution to 0.56 when correction for selection biases and redshift evolution is applied, in the case of the 5$\sigma$ flat priors, which is a 23\% decrease. Similarly, it goes from 0.572 (no-correction) to 0.43  (when correction for selection biases and redshift evolution is applied) in case of the 3$\sigma$ for flat priors, which is a 25\% decrease. Thus, we can conclude that, similarly to the case of the fundamental plane \cite{10.1093/mnras/stac2752,Dainotti2023alternative}, we achieve a reduction of the scatter as when we consider the redshift evolution and selection biases. This, indeed, is the most complete treatment for the use of GRB correlations as cosmological tools. When we consider instead the $b$ parameter the difference between the non-corrected 3$\sigma$ gaussian prior, 0.248, with the respective corrected one, 0.487, is 96\% percentage difference. Respectively, for the 5$\sigma$ gaussian prior the scatter reduction for $b$ goes from 0.25 (non-corrected case) to 0.477 (corrected case) which is a 91\% increase. Last but not least, when we consider the $C_0$ parameter, the difference between the non-corrected 3$\sigma$ flat prior, 48.874, and the corrected one, 25.45, is 48\%; while for the respective 5$\sigma$ case it goes from 41.575 to 25.53, which is a 39\% decrease. This is expected since the error-bars increase when the parameters of the evolution are added, but the most important point is that the overall scatter of the correlation decreases.

\begin{figure}
    \centering
    \includegraphics[width = 0.495\textwidth]{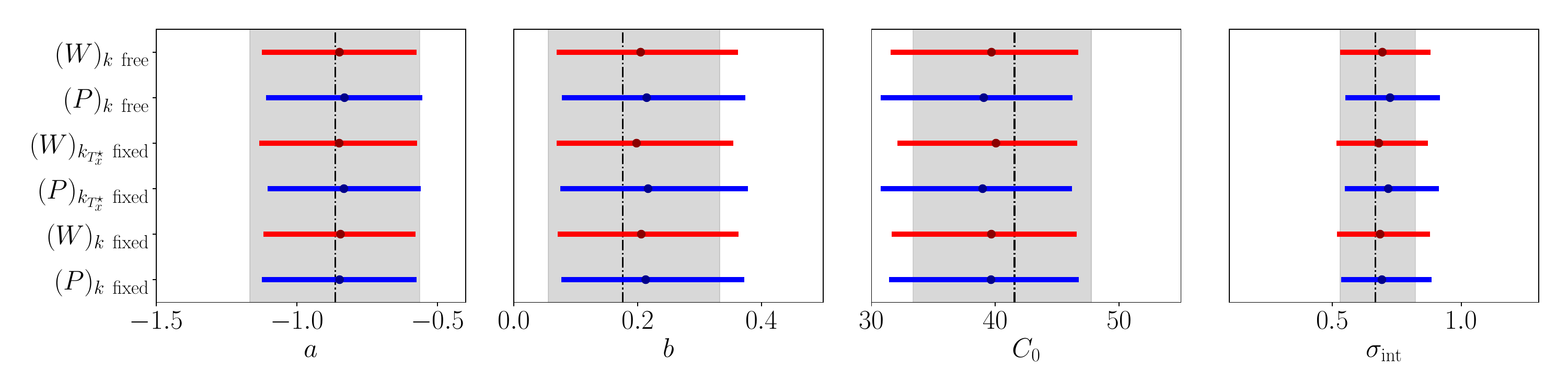}
    \includegraphics[width = 0.495\textwidth]{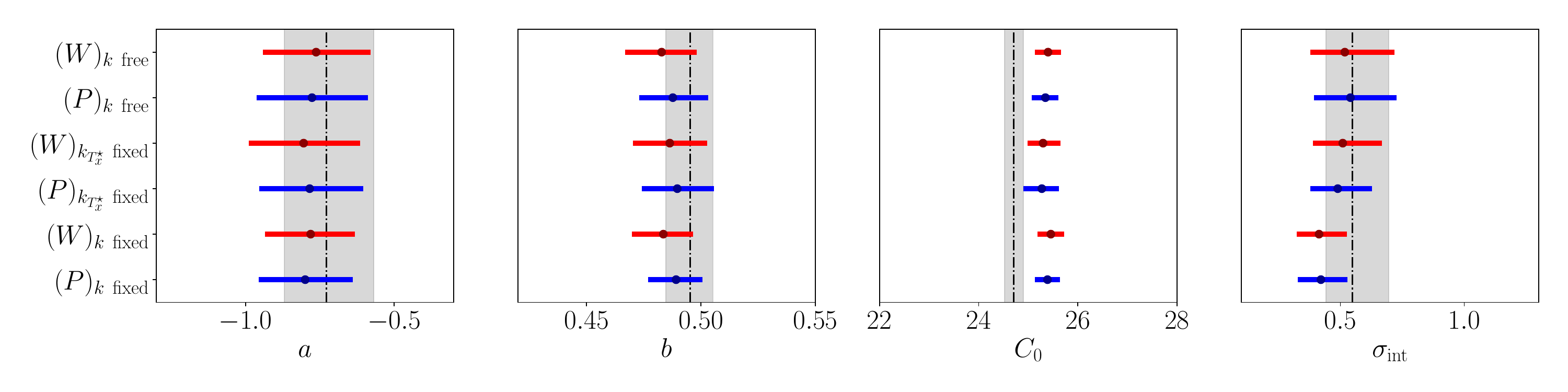} \\
    \includegraphics[width = 0.495\textwidth]{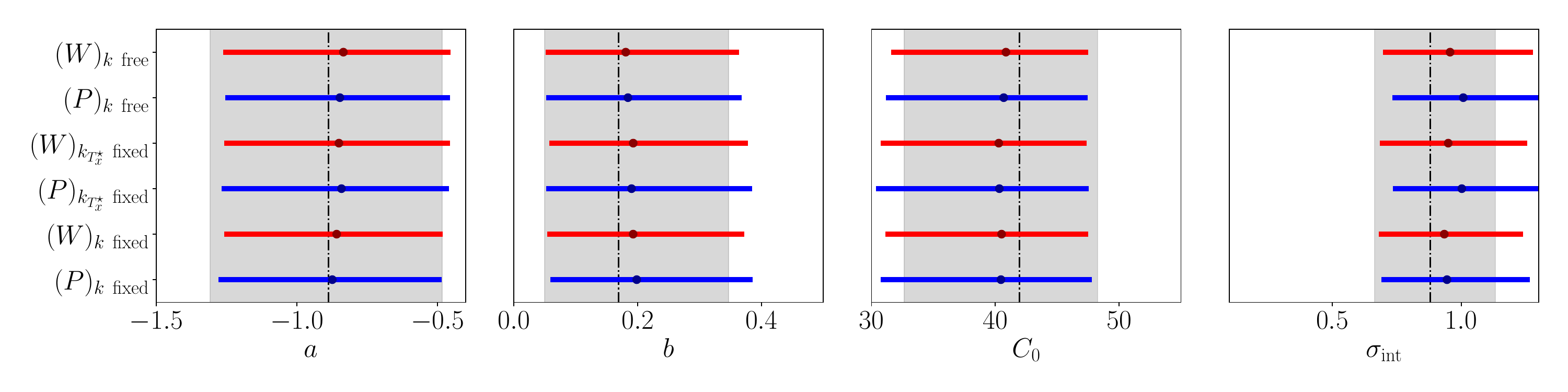}
    \includegraphics[width = 0.495\textwidth]{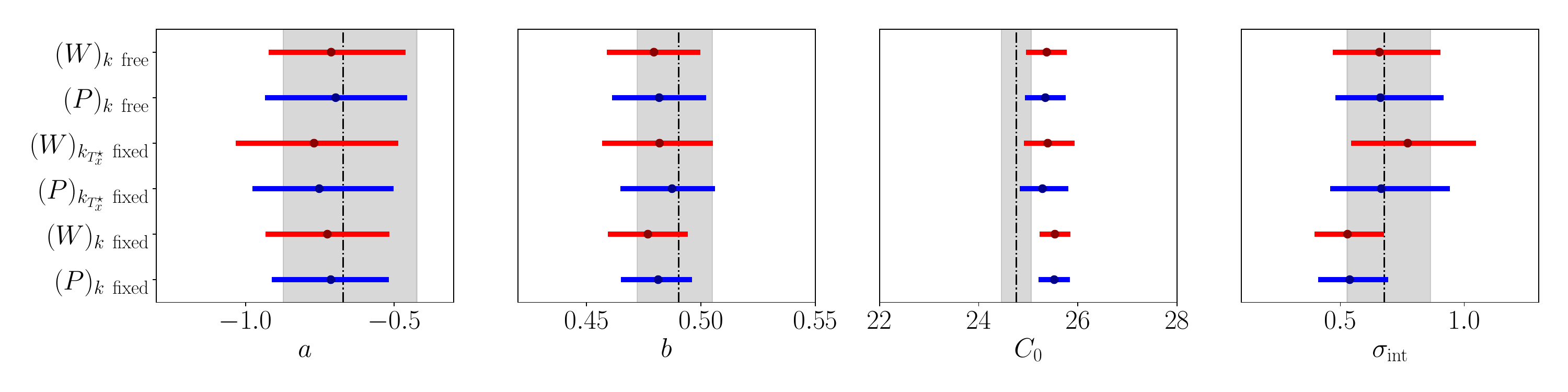}
    \caption{Whisker plots with mean and 68\% C.L. for GRB calibration parameters $a$, $b$, $C_0$ and $\sigma_{\rm int}$ including redshift evolution corrections in the 3D Dainotti relation, sampled assuming Flat (in F.) priors at 3$\sigma$ (top) and $5\sigma$ (bottom) from the Platinum sample (in P.) or Whole sample (in W.), on fixing the GRB nuisance parameters in left panel, with marginalization over the nuisance parameters in right panels. The black dashed line with shaded regions represent the mean and 68\% CL for GRB calibration parameters without introducing redshift evolution correction terms.}
    \label{fig:summary}
\end{figure}

We now study the effect when the redshift evolutionary coefficients are free to vary with the GRB calibration parameters. Here we consider two specific scenarios: (a) We treat ${k_L}_x$ and ${k_L}_{\rm peak}$ as nuisance parameters while  ${k_T^{\star}}_x$ remains fixed (since the latter does not depend on the underlying cosmological framework); (b) All three coefficients are treated simultaneously as nuisance parameters in the MCMC analysis. Notably, the results for calibration parameters $a$, $b$ and $C_0$ are not affected by these choices. In Fig. \ref{fig:kt_fixed} and \ref{fig:all_k_free} we show the corresponding posteriors of the evolutionary coefficients for both the above-mentioned scenarios. The constraints obtained for the redshift evolutionary coefficients, ${k_L}_x$, ${k_T^{\star}}_x$ and ${k_L}_{\rm peak}$, sampled from the Platinum and Whole GRB samples, assuming flat priors in $n$-$\sigma$ progression (with fixed values of GRB nuisance parameters and nuisance marginalization) is mentioned in Tab. \ref{tab:evo_correction_result}. The uncertainties for the latter case are smaller since the fundamental plane fitting analysis for the Whole sample has more constraining power, being the larger data set. We here notice that from our analysis, similar to the analysis performed in \cite{Dainotti:2023pwk} the parameters of the evolution are compatible within 1$\sigma$ for all the evolutionary coefficients showing the reliability of this analysis as well. Furthermore, all these different prior settings provide consistent results for the calibration parameters (see left panels of Figs. \ref{fig:kt_fixed} \& \ref{fig:all_k_free}), even when the redshift evolution corrections are accounted for. Finally, for the main parameters of interest, we show their constraints with a whisker plot in Fig. \ref{fig:summary}, to compare the results throughout all the different cases described above, with and without selection bias effects. 

However, some differences are observed in the right and middle panels for Fig. \ref{fig:kt_fixed} \& \ref{fig:all_k_free}, when we marginalize over the nuisance parameters defining the GRB fundamental plane. With this scenario, the redshift evolution correction terms are tightly constrained compared to the respective cases on keeping the nuisance parameters fixed during the MCMC analysis. The results from 3$\sigma$ nuisance priors for the Whole sample vs the Platinum sample, show a tension in the parameter spaces for the evolutionary coefficients, particularly ${k_L}_x$ and ${k_L}_{\rm peak}$. This tension is somewhat relaxed when $5\sigma$ priors are considered for marginalization over the nuisance parameters. Notably, the constraints on ${k_{T}}^{*}_{x}$ are not significantly affected during this analysis, as expected, being related to a measure of a characteristic time scale for the end of the plateau emission, and not of the underlying cosmology.

\section{Comparison with Gaussian Processes}  \label{sec:comparison_discussion}

\begin{figure}[t]
    \centering
    \includegraphics[width = 0.565\textwidth, height=0.13\textheight]{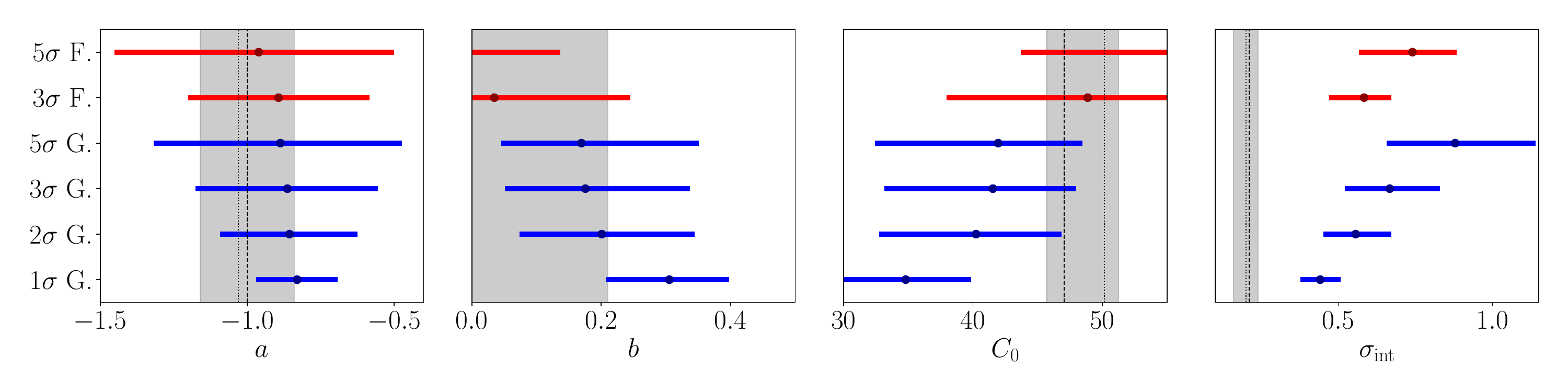}
    \includegraphics[width = 0.425\textwidth, height=0.125\textheight]{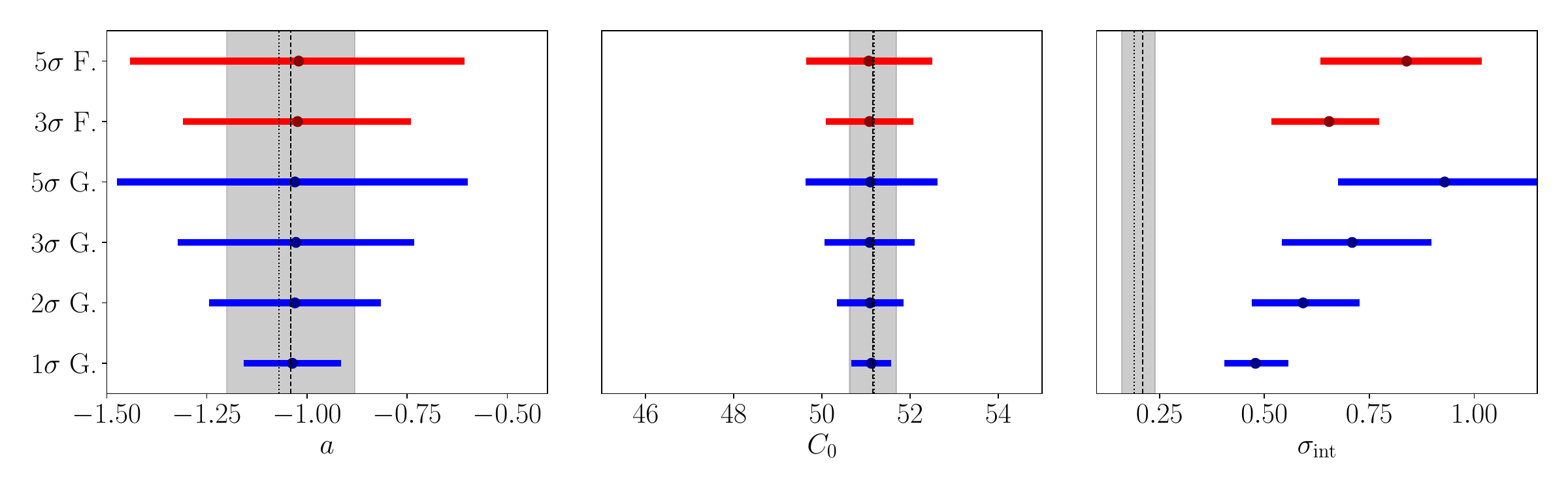}\\
    ~~~~~~~~~~~~~~~~~~~~~~~~~~~~~~~~~~~~~~~~~~~~~~~ (a) \hfill (b) ~~~~~~~~~~~~~~~~~~~~~~~~~~~~~~~~~~~~
\caption{Comparison between the constraints obtained on the GRB calibration parameters employing neural networks (shown via whisker plots) vs Gaussian processes (shown with shaded regions) from Ref.  \cite{Favale:2024lgp} employing the (a) 3D Dainotti relation and (b) 2D Dainotti relation, respectively.}
    \label{fig:gp_compare}
\end{figure}

This section provides a detailed comparison of the results obtained from Artificial Neural Networks (ANN) and Gaussian Processes (GP) in the context of four calibration parameters relevant to GRBs: $a$, $b$, $C_0$, and $\sigma_{\rm int}$. We first compare the 1D posteriors and 2D contours at 68\% and 95\% C.L.s for the 3D and 2D Dainotti relations (see Fig.~\ref{fig:3d_cal_mcmc} \& \ref{fig:2d_cal_mcmc} in our work and Fig.~3 and 8 of Ref. \cite{Favale:2024lgp}). Overall, the results are consistent, especially at the 3$\sigma$ and 5$\sigma$ levels. The larger differences in the 1$\sigma$ case, for both ANNs and GPs, could likely arise because the walkers explore only a limited region of the parameter space around the mean values of the FP. We further compare our findings with the summary in Table 2 of Ref.~\cite{Favale:2024lgp}, illustrated with a whisker plot in Fig.~\ref{fig:gp_compare}. 
\begin{itemize}[left=0pt]
\item $\boldsymbol{a}$: For the 3D Dainotti relation, ANN estimates $a$ as $ -0.961_{-0.480}^{+0.451} $. This substantial uncertainty reflects the model's adaptability to various data scenarios, but it raises concerns about overfitting, especially in small datasets. GP, in contrast, provides a more constrained estimate of $-0.98 \pm 0.16$, which results in a narrower range of approximately $-1.14$ to $-0.82$. The GP's tighter CLs highlight its effectiveness in utilizing prior knowledge, leading to clearer and more precise parameter estimates. For the 2D relation, ANN yields $a = -1.021_{-0.413}^{+0.405} $, again demonstrating flexibility but with similar concerns regarding potential overfitting. Likewise, GP yields $-1.01 \pm 0.17$, reinforcing its strength in leveraging prior information for enhanced precision.
\item $\boldsymbol{b}$: Both methods struggle to constrain $b$, yielding only upper limits: $<0.250$ (ANN) and $<0.21$ (GP) for the 3D relation, indicating a common limitation. This consistency across both methods suggests that $b$ remains poorly constrained in the context of GRB data, reflecting a common difficulty in accurately modeling $b$. Similarly, for the 2D relation, we find $b <0.218$ (ANN), while $<0.21$ (GP). Thus, $b$ presents challenges for both methods, yielding only upper limits, indicating that further data or modeling strategies may be necessary to improve constraints.
\item $\boldsymbol{C_0}$: The calibration parameter $C_0$ demonstrates different strengths for each method. In the 3D case, ANN estimates $C_0 = 25.807 \pm 8.978 $, indicating a broader variability that may complicate accuracy. This adaptability is beneficial in modeling diverse data scenarios, yet it can lead to challenges in interpretation. Conversely, GP provides a more definitive estimate of $46.96^{+4.25}_{-1.38}$, showcasing its reliability and stability. In the 2D case, ANN reports $ 51.063_{-1.343}^{+1.365} $, showing improved precision, while GP offers $51.11 \pm 0.54$. The narrower C.L. for GP highlight its capacity to incorporate prior information effectively, resulting in a clearer understanding of this parameter. 
\item $\boldsymbol{\sigma_{\rm int}}$: The intrinsic scatter, $\sigma_{\rm int}$, serves as a critical measure of the variability in the data. For the 3D Dainotti relation, ANN estimates $ 0.337 \pm 0.070 $, indicating significant variability. While this reflects ANN's ability to model complex behaviors, it raises concerns about overfitting, particularly in smaller datasets. In contrast, GP provides a more conservative estimate of $0.22^{+0.03}_{-0.05}$, suggesting a more cautious approach, yet limited ability to capture intricate patterns. For the 2D relation, ANN gives $ 0.839_{-0.197}^{+0.171} $, while GP gives $0.21^{+0.03}_{-0.05}$, highlighting its tighter CL and conservative assessment.
\end{itemize}
In summary, ANN generally provides broader ranges and higher estimates for $a$ and $\sigma_{\rm int}$, reflecting its flexibility in accommodating complex data structures. However, this flexibility can lead to challenges with overfitting and the necessity for extensive hyperparameter tuning. Conversely, the GP approach offers tighter estimates and greater interpretability, particularly for parameters $C_0$ and $\sigma_{\rm int}$. However, its reliance on prior information may limit its adaptability to more intricate patterns present in the data. Both methods face difficulties in accurately estimating the parameter $b$, suggesting that improvements in data quality or modeling techniques may be necessary. This detailed analysis underscores the complementary nature of ANN and GP, emphasizing that the choice between these approaches should be context-dependent, balancing the need for flexibility in modeling with the requirement for precision and interpretability in GRB research.

\section{Cosmological Parameter Inference} \label{sec:cosmo}

Building on the stability of the results established in the previous sections, we now focus on the estimation of cosmological parameters, utilizing the calibrated relations to extract meaningful constraints on the underlying cosmological models. The calibrated GRBs, will help to extend the distance ladder beyond SNia and BAO redshifts. Thus, we employ the calibration results obtained by making use of the GRB data at z $\leq$ 2 to obtain luminosity distances from the remaining GRBs contained in the Platinum sample and cover the region up to $z = 5$ building the extended distance ladder. 
Herein, we present a two-step analysis, where the extended distance ladder obtained with our baseline calibration without considering evolution effects and compare it with the one obtained when the latter are also taken into account. To quantify deviations from the ANN-based reconstructed Pantheon+ distances, we compute the bias for each of the calibrated GRB distances at each redshift, $\beta = \log D_{\rm GRB}^L - \log D_{\rm ANN}^L$. To track this trend more precisely, we bin the redshift range (50 data points) into 10 equi-populated bins. The weighted mean and its uncertainty for each bin are computed as follows:
\begin{equation}
\tilde{\beta}_k = \frac{\sum_{i,j \in e_k} \beta_i \omega_{ij}}{\sum_{i,j \in e_k} \omega_{ij}}, \quad \sigma_k = \sqrt{\frac{1}{\sum_{i,j \in e_k} \omega_{ij}}}
\end{equation}
where $\omega_{ij} = (C^{-1})_{ij}$ is the inverse covariance.

\begin{figure}[t]
    \centering
    \begin{minipage}[b]{0.5\textwidth}
        \centering
        \includegraphics[width=\textwidth]{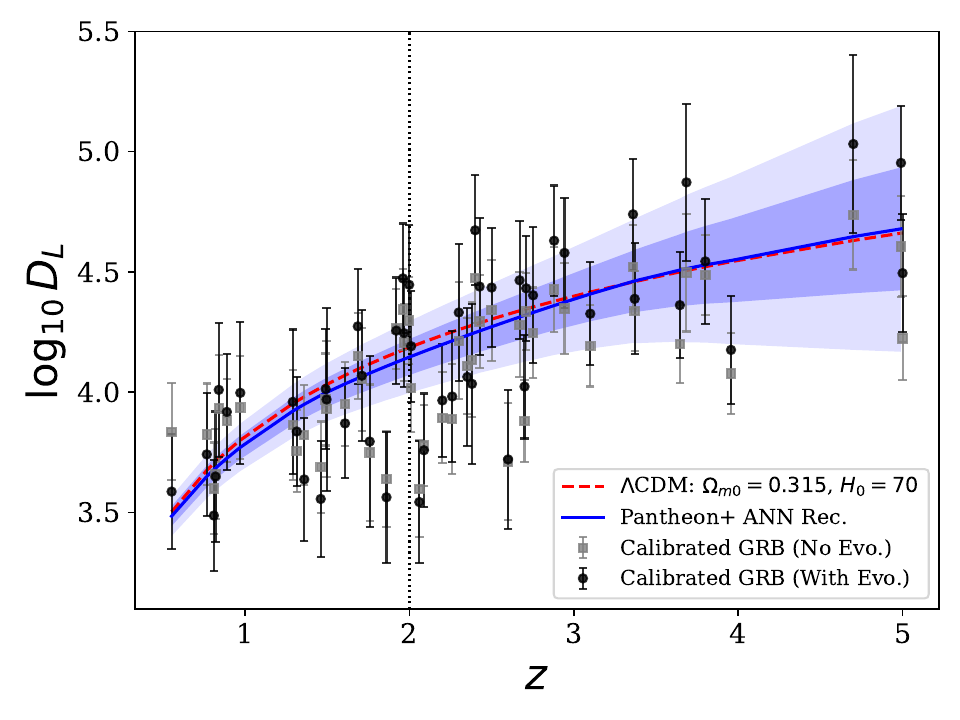}
    \end{minipage}%
    \begin{minipage}[b]{0.5\textwidth}
        \centering
        \includegraphics[width=\textwidth, height=0.4\textwidth]{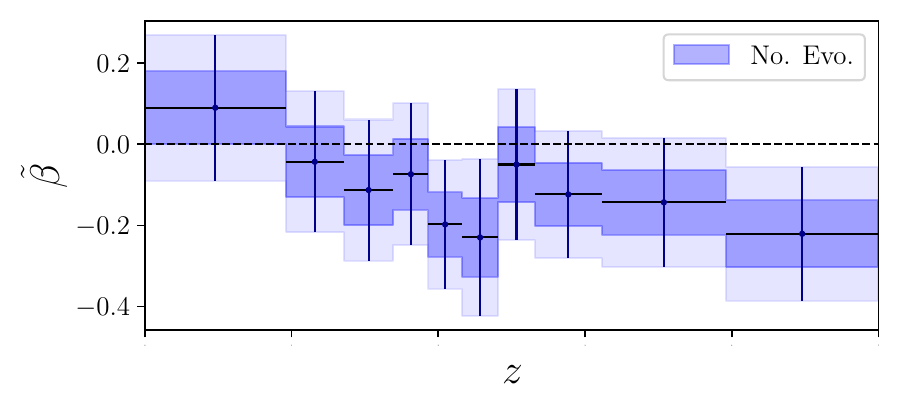}\\[-0.8cm]
        \includegraphics[width=\textwidth, height=0.4\textwidth]{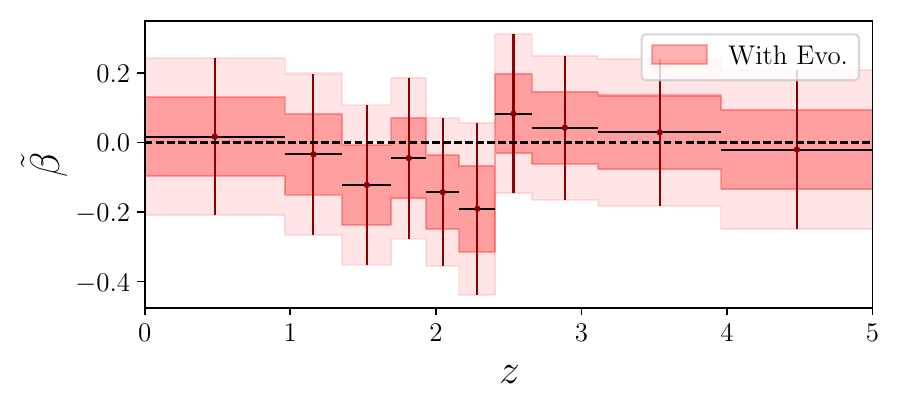}
    \end{minipage}
    \caption{Comparison of calibration of the distance luminosity (extrapolated to higher redshifts) with and without evolution.}
    \label{fig:dl_calib}
\end{figure}

In the calibrated GRB data points of $\log_{10} D_L(z)$ without evolution effects, a negative $\beta$ trend is evident for $z > 2$, consistent with previous findings (e.g., Postnikov et al.\cite{postnikov2014}, Favale et. al.\cite{Favale:2024lgp}). However, in the case where the redshift evolution effects are considered, $\beta=0$ is always included within 2$\sigma$.  Thus, on comparing the upper and lower panels, we can infer that accounting for redshift evolution corrections is extremely important to decrease the bias in the region $z > 2$.  

When evolution effects are ignored, a clear trend towards smaller luminosity distances is observed (see left panel of Fig. \ref{fig:dl_calib}). We demonstrate that accounting for evolutionary effects effectively mitigates this trend, particularly when considering the correlations between various GRB luminosity distances. It is well-established that the physical properties of astrophysical objects at high redshifts are more influenced by the Malmquist bias \cite{1922MeLuF.100....1M}, due to the challenges of detecting faint events at greater distances. This highlights the importance of properly correcting for biases and redshift evolution effects when using high-redshift probes for cosmological studies.

We now proceed to derive the cosmological parameters, $H_0$ and $\Omega_{m}$, from these calibrated GRB distances, assuming a flat-$\Lambda$CDM model. To this end, we explore four cosmological priors$-$
\begin{itemize}
    \item \textbf{Case (1)}: 5$\sigma$ for the fiducial values of the Planck priors in both parameters; 
    \item \textbf{Case (2)}: 5$\sigma$ for the fiducial values for the Pantheon+ \& SH0ES priors;
    \item \textbf{Case (3)}: 1$\sigma$ for $\Omega_{m}$ taking from Planck and we leave flat priors for $10 < H_0 < 200$, so completely uninformative;
    \item \textbf{Case (4)}: 1$\sigma$ for $\Omega_m$ taking from Pantheon+ \& SH0ES and we leave the same flat priors for $10 < H_0 < 200$.
\end{itemize}

\begin{figure}[t]
    \centering
    \includegraphics[width = 0.4\textwidth]{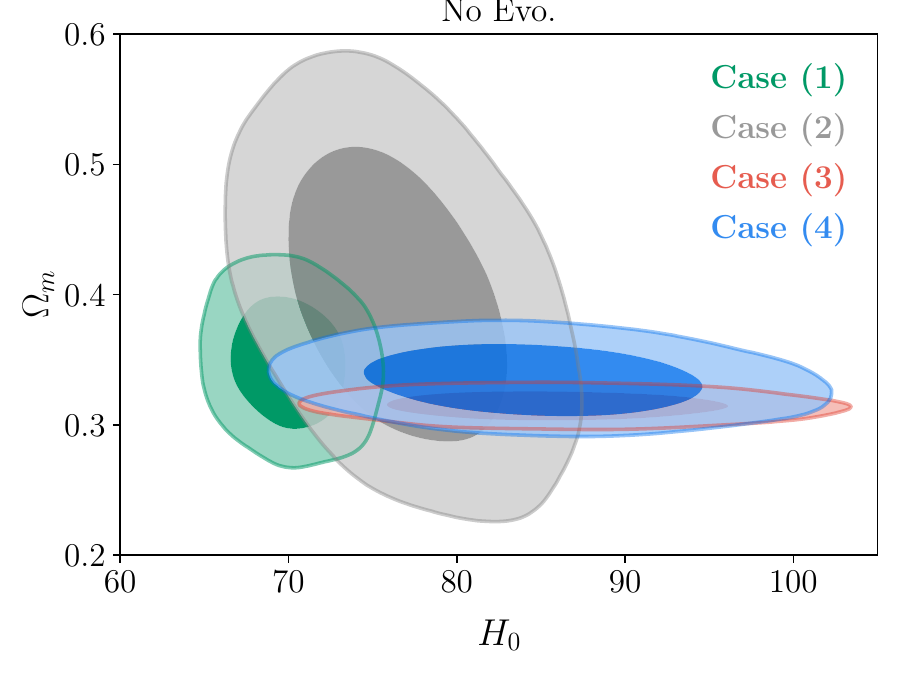}
    \includegraphics[width = 0.4\textwidth]{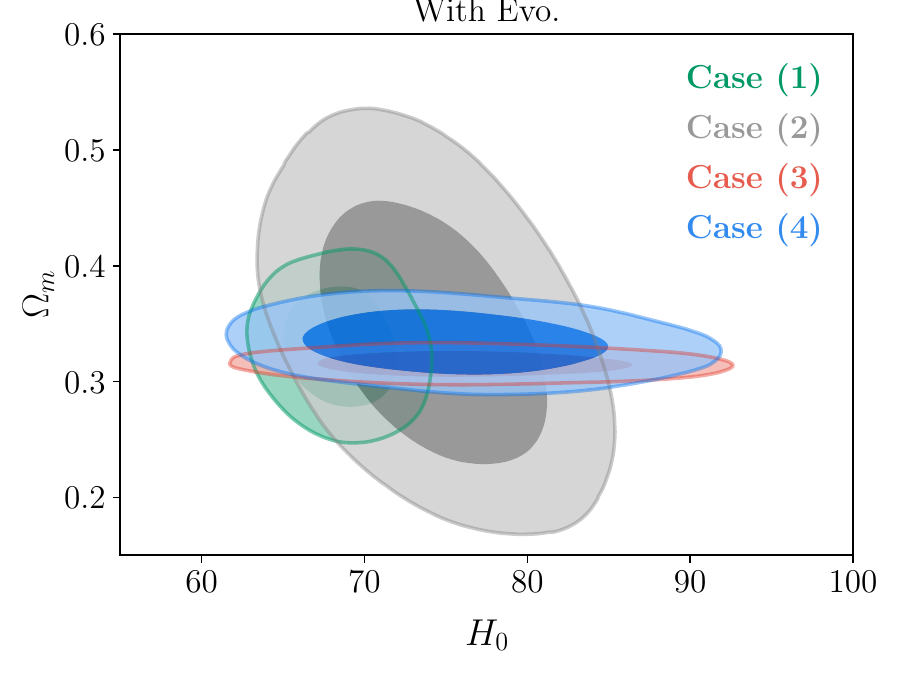}
    \caption{Two dimensional confidence contours showing the constraints on the $H_0-\Omega_m$ parameter space with the calibrated GRB data, excluding redshift evolution correction (i.e., No Evo. in left panel) and including redshift evolution correction (i.e., With Evo. right panel).}
    \label{fig:cosmo}
\end{figure}

\begin{table}[t]
{\renewcommand{\arraystretch}{1.5} \setlength{\tabcolsep}{8 pt} \centering   
\begin{tabular} { l  c c c c}
\noalign{\vskip 3pt}\hline\noalign{\vskip 1.5pt}\hline\noalign{\vskip 5pt}
 \multicolumn{1}{c}{\bf } &  \multicolumn{1}{c}{\bf Case (1)} &  \multicolumn{1}{c}{\bf Case (2)} &  \multicolumn{1}{c}{\bf Case (3)} &  \multicolumn{1}{c}{\bf Case (4)}\\
\noalign{\vskip 3pt}\cline{2-5}\noalign{\vskip 3pt}
{\bf Parameter} &  \multicolumn{4}{c}{\bf No Evo.}\\
\hline
{\boldmath$H_0$ [in km Mpc$^{-1}$ s$^{-1}$]      } & $70.0\pm 2.3               $ & $76.5^{+4.0}_{-4.5}        $ & $86.0^{+5.9}_{-7.2}        $ & $84.8^{+6.2}_{-6.9}        $\\

{\boldmath$\Omega_m$     } & $0.349\pm 0.034            $ & $0.400\pm 0.074            $ & $0.3148\pm 0.0073          $ & $0.335\pm 0.018            $\\
\hline
\textbf{Parameter} &  \multicolumn{4}{c}{\bf With Evo.}\\
\hline
{\boldmath$H_0$ [in km Mpc$^{-1}$ s$^{-1}$]         } & $68.6\pm 2.3               $ & $74.2\pm 4.5               $ & $76.6^{+5.7}_{-6.8}        $ & $75.7^{+5.5}_{-6.7}        $\\

{\boldmath$\Omega_m$     } & $0.330\pm 0.034            $ & $0.343^{+0.071}_{-0.080}   $ & $0.3151\pm 0.0073          $ & $0.334\pm 0.018            $\\
\hline
\end{tabular}
}
\caption{Constraints on the cosmological parameters $H_0$, $\Omega_m$, obtained from the calibrated GRBs assuming a flat-$\Lambda$CDM cosmological model. We report the results where we take into account calibration of GRBs both excluding (No Evo.) and including (With Evo.) redshift evolution corrections. }
\label{tab:cosmo}
\end{table}

The result of our analysis is shown in Table \ref{tab:cosmo}. The two-dimensional confidence contours for $H_0-\Omega_m$ is shown in Fig. \ref{fig:cosmo}. Our findings clearly demonstrate that properly accounting for redshift evolution in the calibration of GRB distances has a significant quantitative impact on the inferred cosmological parameters. In particular, including evolution corrections systematically lowers the best-fit values of $H_0$ and $\Omega_m$, and leads to more consistent results across different prior choices.

Without applying redshift evolution corrections, the inferred $H_0$ values are systematically higher. Under weakly informative priors (Cases 3 and 4), $H_0$ exceeds $80\,\mathrm{km\,s^{-1}\,Mpc^{-1}}$, reaching $86.0^{+5.9}_{-7.2}$ and $84.8^{+6.2}_{-6.9}\,\mathrm{km\,s^{-1}\,Mpc^{-1}}$, respectively. After correcting for evolution, these values drop by about $10\,\mathrm{km\,s^{-1}\,Mpc^{-1}}$, down to $76.6^{+5.7}_{-6.8}$ and $75.7^{+5.5}_{-6.7}\,\mathrm{km\,s^{-1}\,Mpc^{-1}}$. Similarly, for tighter priors (Cases 1 and 2), $H_0$ decreases from $70.0\pm2.3$ and $76.5^{+4.0}_{-4.5}$ to $68.6\pm2.3$ and $74.2\pm4.5\,\mathrm{km\,s^{-1}\,Mpc^{-1}}$, respectively, when evolution corrections are included.

A similar behavior is observed for the matter density parameter $\Omega_m$. Without evolution corrections, $\Omega_m$ is consistently overestimated. For instance, in Case 1, it shifts from $0.349\pm0.034$ (No Evo.) to $0.330\pm0.034$ (With Evo.), and in Case 2 from $0.400\pm0.074$ to $0.343^{+0.071}_{-0.080}$.

Moreover, the influence of different prior choices is evident: tighter priors on $\Omega_m$ yield stronger constraints on $H_0$, while flat priors permit a broader range of values, as expected. Overall, these results highlight the critical importance of properly modeling redshift evolution effects when using GRBs as cosmological probes. Ignoring evolution leads to systematic biases, whereas correcting for it enables more reliable, unbiased, and internally consistent cosmological constraints, especially at high redshift.

\section{Discussion and Conclusions} \label{sec:conclusion}

In this work, we selected a high-quality GRB sample following established criteria, focusing on the Platinum dataset detailed in \cite{Dainotti:2020azn}. Each GRB light curve was modeled with a parametric function describing both the prompt gamma-ray and early X-ray afterglow phases, using two sets of four parameters: $T_i$, $F_i$, $\alpha_i$, and $t_i$. This fitting yielded a subsample of 50 GRBs with reliably determined luminosity and plateau properties. We derived the peak prompt luminosity $(L_{\rm peak})$ and plateau X-ray luminosity $(L_X)$ using a standard luminosity-distance relation that includes cosmological parameters and a K-correction. Building on the well-known 2D Dainotti relation between $L_X$ and plateau duration $T^{*}_X$, we further explored the 3D Dainotti relation involving $L_{\rm peak}$, $L_X$, and $T^{*}_X$, which significantly reduces scatter and enhances GRBs' role as standard candles. The 3D relation, with a slope close to the magnetar model expectation of $-1$, reinforces its intrinsic physical origin.

To calibrate these relations without assuming a cosmological model, we developed a two-stage ANN framework. First, an ANN was trained on the Pantheon+ SNIa dataset to reconstruct the Hubble diagram, and a second ANN was then used to constrain the GRB systematics. Compared to previous methods based on Gaussian processes \cite{Favale:2024lgp}, our approach mitigates concerns about overfitting and kernel dependence. Calibration was performed using MCMC techniques, examining both flat and Gaussian prior choices for the parameters $a$, $b$, $C_0$, and $\sigma_{\rm int}$. Results are generally stable across prior choices, with minor differences observed under tighter ($1\sigma$) constraints, particularly affecting $b$ and $C_0$. Our findings reveal that $\sigma_{\rm int}$ shows notable dependence on the systematic parameters when accounting for redshift evolution. Specifically, the intrinsic scatter $\sigma_{\rm int}$ is reduced, $a$ prefers slightly higher values than $-1$, and $b$ exhibits a marginal tendency towards positive values, all of which could impact cosmological inferences. Importantly, a strong degeneracy between $C_0$ and $b$ emerges, emphasizing the critical role of redshift evolution corrections.

Extending the calibrated GRB sample to higher redshifts enabled us to infer cosmological parameters $H_0$ and $\Omega_m$ under a flat-$\Lambda$CDM framework. Accounting for redshift evolution significantly improves the consistency of cosmological constraints across different prior assumptions, systematically lowering both $H_0$ and $\Omega_m$ estimates. In particular, neglecting evolution leads to higher $H_0$ values by approximately $10\,\mathrm{km\,s^{-1}\,Mpc^{-1}}$, and consistently overestimates $\Omega_m$, especially when non-informative priors are used. These results underscore the necessity of model-independent calibration methods for extending the cosmic distance ladder. GRBs offer a promising probe at high redshifts, but their variability and uncertain progenitors demand careful standardization. Model-independent techniques, like the one employed here, are essential for minimizing biases, reducing systematic uncertainties, and enabling GRBs to contribute robustly to precision cosmology.

\begin{acknowledgments}
This article is also based upon work from COST Action CA21136 Addressing observational tensions in cosmology with systematics and fundamental physics (CosmoVerse) supported by COST (European Cooperation in Science and Technology). PM would like to acknowledge funding from ANRF SERB, Govt. of India under the National Post Doctoral Fellowship (File no. PDF/2023/001986). PM acknowledges the use of High Performance Computing facility Pegasus at IUCAA, Pune, India. JLS and JM would also like to acknowledge funding from ``Xjenza Malta'' as part of the ``FUSION R\&I: Research Excellence Programme'' REP-2023-019 (CosmoLearn) Project, and the ``Net4Tensions'' as part of the ``Research Networking Scheme''. K.F.D. was supported by the PNRR-III-C9-2022–I9 call, with project number 760016/27.01.2023. M.G.D. acknowledges the support of the DoS and by JSPS Grant-in-Aid for Scientific Research (KAKENHI) (A), Grant Number JP25H00675.
\end{acknowledgments}

\bibliography{references}

\end{document}